\documentclass[journal,twoside,web]{ieeecolor}

\usepackage{generic}
\usepackage{cite}
\usepackage{algorithmic}
\usepackage{graphicx}
\usepackage{amsmath} 
\usepackage{amssymb} 
\usepackage{amsfonts,mathtools,bbm,verbatim}
\usepackage{enumerate}
\usepackage{url}  
\usepackage{bbm}
\usepackage{etoolbox}
\usepackage{mathrsfs}
\usepackage[utf8]{inputenc}
\usepackage{hyperref}
\hypersetup{colorlinks = true, linkcolor = blue, urlcolor = blue}

\usepackage{breakurl}
\allowdisplaybreaks
\usepackage{balance}

\def\BibTeX{{\rm B\kern-.05em{\sc i\kern-.025em b}\kern-.08em
    T\kern-.1667em\lower.7ex\hbox{E}\kern-.125emX}}
\def\R{{\mathbb{R}}}
\def\Z{{\mathbb{Z}}}
\def\P{{\mathbb{P}}}
\def\E{{\mathbb{E}}}
\def\I{{\mathbbm{1}}}
\def\G{{\mathcal{G}}}
\def\N{{\mathcal{N}}}

\def\F{{\mathcal{F}}}
\def\H{{\mathcal{H}}}
\def\Ber{{\mathsf{Ber}}}
\def\bin{{\mathsf{bin}}}

\newtheorem{Theorem}{Theorem}
\newtheorem{Lemma}{Lemma}
\newtheorem{Proposition}{Proposition}
\newtheorem{Corollary}{Corollary}

\newtheorem{Definition}{Definition}

\newcommand{\secref}[1]{Section~\ref{#1}}
\newcommand{\appref}[1]{Appendix~\ref{#1}}
\newcommand{\thmref}[1]{Theorem~\ref{#1}}
\newcommand{\propref}[1]{Proposition~\ref{#1}}
\newcommand{\lemref}[1]{Lemma~\ref{#1}}

\begin{document}

\title{Inference in Opinion Dynamics\\under Social Pressure} 

\author{Ali Jadbabaie, Anuran Makur, Elchanan Mossel, and Rabih Salhab
\thanks{The author ordering is alphabetical.}%
\thanks{A. Jadbabaie is with the Department of Civil and Environmental Engineering, Massachusetts Institute of Technology, Cambridge, MA 02139, USA (e-mail: jadbabai@mit.edu).}%
\thanks{A. Makur is with the Department of Computer Science and the Elmore Family School of Electrical and Computer Engineering, Purdue University, West Lafayette, IN 47907, USA (e-mail: amakur@purdue.edu).}%
\thanks{E. Mossel is with the Department of Mathematics, Massachusetts Institute of Technology, Cambridge, MA 02139, USA (e-mail: elmos@mit.edu).}%
\thanks{R. Salhab is with Amazon, USA (email: rabihisalhab@gmail.com).}%
\thanks{This work was supported in part by the ARO MURI grant W911NF-19-1-0217, a Vannevar Bush Fellowship N00014-16-1-3163 from OUSD, a Vannevar Bush Faculty Fellowship ONR-N00014-20-1-2826, a Simons Investigator award (622132), and an NSF award CCF-1918421.}}%

\maketitle

\begin{abstract}
We introduce a new opinion dynamics model where a group of agents holds two kinds of opinions: inherent and declared. Each agent's inherent opinion is fixed and unobservable by the other agents. At each time step, agents broadcast their declared opinions on a social network, which are governed by the agents' inherent opinions and social pressure. In particular, we assume that agents may declare opinions that are not aligned with their inherent opinions to conform with their neighbors. This raises the natural question: Can we estimate the agents' inherent opinions from observations of declared opinions? For example, agents' inherent opinions may represent their true political alliances (Democrat or Republican), while their declared opinions may model the political inclinations of tweets on social media. In this context, we may seek to predict the election results by observing voters' tweets, which do not necessarily reflect their political support due to social pressure. We analyze this question in the special case where the underlying social network is a complete graph. We prove that, as long as the population does not include large majorities, estimation of aggregate and individual inherent opinions is possible. On the other hand, large majorities force minorities to lie over time, which makes asymptotic estimation impossible.
\end{abstract}


\section{Introduction}

Since its inception in the seminal works of French Jr. \cite{French1956} and DeGroot \cite{DeGroot1974}, the mathematical theory of opinion dynamics has evolved into a very pertinent area of research today. This theory tries to understand and analyze the formation and evolution of opinions in groups of individuals pertaining to some topic of common interest, e.g., political support, climate change, etc. Indeed, opinion dynamics models shed light on several important phenomena, such as, consensus (convergence to a common opinion), rate of convergence, and social learning (i.e., whether the consensus opinion matches the outcome of a centralized learning problem). The vast majority of models that have been studied in the literature share two features. Firstly, the agents are reliable in that they share their opinions truthfully. Secondly, the agents update their opinions over time according to a predefined rule. 

However, there is increasing evidence that people do not always share what they truly believe. For example, a recent poll showed that $62\%$ of Americans do not share their true political views out of fear of offending others \cite{poll2020}. Moreover, over the last few decades, psychologists have designed experiments to demonstrate that individuals often conform with the opinions of their peers due to a desire to be accepted by the group, even if the group's opinion does not align with their true beliefs \cite{cialdini2004social,crutchfield1955conformity,insko1983conformity,asch1951effects,asch1955opinions}. This suggests that individuals have two kinds of opinions: \emph{inherent} and \emph{declared}. The former is the individual's true belief, which is unobservable by other individuals. The latter is what individuals declare in public. More importantly, individuals may lie and declare opinions that are not aligned with their inherent opinion to conform with others. This raises the natural question: \emph{Can we estimate individuals’ inherent opinions from observations of their declared opinions?} 

Answering this question is important in many contexts. For instance, it has been shown that conformity can be catastrophic for legal and political institutions as it deprives society of important information \cite{hibbing2002stealth, kalven1966american, sunstein2002conformity}. In particular, Sunstein argued that a Democratic judge sitting with two Republican judges is more likely to reverse an environmental decision at the behest of an industry challenger \cite{sunstein2002conformity}. Thus, estimating the amount of compliance under social pressure helps in assessing how freely deliberations are held, which is crucial for the functioning of deliberative democracies and trial juries. 

In this paper, we introduce a new opinion dynamics model that involves a group of connected agents with both inherent and declared binary opinions. For example, an agent's inherent opinion might represent their true political stance (Republican or Democrat), and their declared opinions might convey the political inclinations of their tweets or posts on social media. We assume that an agent's inherent opinion is fixed and unobservable by the other agents. On the other hand, an agent's declared opinions are updated over time and broadcast over a social network. Furthermore, each agent's truth probability, i.e., the probability with which the agent declares their inherent opinion, follows the well-known Bradley-Terry-Luce model \cite{Zermelo1929,BradleyTerry1952,Luce1959,McFadden1973}; so, it is proportional to the number of times the agent observes broadcast opinions that agree with their inherent opinion. In other words, an agent gains more confidence to tell the truth when they observe more declared opinions aligned with their inherent one. The goal of this work is to construct an estimator that infers the agents' inherent opinions based on observations of their (unreliable) declared ones.

\subsection{Contributions}
\label{Contributions}

This paper makes the following main contributions:
\begin{enumerate}
\item We introduce a new \emph{interacting P\'{o}lya urn model} of opinion dynamics where agents hold inherent and declared binary opinions. The former are not observable, while the latter are influenced by social pressure and broadcast on a social network. Naturally, like other mathematical models of opinion dynamics, our model does not capture many of the nuanced cognitive and sociological aspects of the real-world opinion dynamics. So, we outline several limitations of our model in \secref{sec:model}. Despite these shortcomings, we feel that our model begins to rigorously describe a new phenomenon where agents possess two kinds of opinions.
\item We analyze the setting where the social network is a complete graph (for analytical tractability). We propose a simple estimator that determines whether there is a large majority of inherent opinions in the population, and when there is no large majority, it asymptotically learns the proportion of agents that have a certain inherent opinion by observing declared opinions over time. The existence of a large majority, however, forces the minority to lie, which makes asymptotic estimation impossible. In this case, our estimator produces a lower bound on the majority's size. 
\item When the population does not have a large majority, we also propose an estimator that learns agents' individual inherent opinions by observing their declared ones.
\end{enumerate}

\subsection{Related Literature}

Since the seminal works of French Jr. \cite{French1956} and DeGroot \cite{DeGroot1974}, a myriad of opinion dynamics models have been introduced in the literature. We do not attempt to survey these models here, and instead only discuss the most important and relevant ones. Perhaps the most renowned (non-Bayesian) opinion dynamics model is the \emph{DeGroot learning model} \cite{DeGroot1974}. In this model, agents update their beliefs about a topic of interest using a weighted average of their neighbors' beliefs. By appealing to the theory of Markov chains, it can be shown that consensus can be reached in such a model. Friedkin and Johnsen extended the DeGroot model to address the fact that people often disagree in the real world \cite{FriedkinJohnsen1990}. They argued that this disagreement was a result of individuals trying to adhere to their initial opinions as much as possible. Specifically, they showed that if agents updated their opinions as a convex combination of their initial opinions and a weighted average of their neighbors' opinions, disagreement could occur. Peng \emph{et al.} extended the DeGroot and Friedkin and Johnsen's models to analyze self-appraisal and social power in opinion dynamics models \cite{jia2015opinion} (also see \cite{jia2013dynamics}).
Moreover, Ye \emph{et al.} also considered an extension of Friedkin and Johnsen's model where they assumed that agents have private and expressed opinions, and both were updated linearly over time \cite{ye2019influence}. These authors studied the asymptotics of the opinion processes, and analyzed the limiting  discrepancy between expressed and private opinions as well as disagreement between private opinions. We refer readers to \cite{ye2019thesis} and the references therein for related work on social power in opinion dynamics models that are variants of the DeGroot and Friedkin and Johnsen's models.
Recently, in a slightly different vein, Gaitonde \emph{et al.} have built on Friedkin and Johnsen's model to introduce a game where an attacker tries to increase opinion discord in a population and a defender intervenes against such attacks \cite{GaitondeKleinbergTardos2020}.

One of the first stochastic opinion dynamics models, known as the \emph{voter model}, was introduced by Holley and Liggett in \cite{holley1975ergodic}. This model considers a group of connected agents with binary opinions. At different instants of time, an agent is picked at random and decides whether to flip its opinion or not. The transition rate of the flipping probability is a function of the agent's and its neighbors' current opinions. Different instances of voter models in the literature correspond to different transition rate functions. For example, linear voter models assume that the transition rate is a convex combination of the neighbors' opinions \cite[Chapter 5]{liggett2012interacting}. The theory of voter models tries to understand the relationship between transition rate functions and possible asymptotic behaviors, e.g., consensus, clustering, coexistence, etc. Other interacting particle systems have also been utilized as stochastic models of opinion dynamics. For instance, Montanari and Saberi proposed a model where the probability of adopting a certain opinion is a softmax function of the number of neighbors that currently hold this opinion, i.e., the adoption probability obeys \emph{Glauber dynamics} for the \emph{Ising model} \cite{MontanariSaberi2010}. 

In a different vein, Hayhoe \emph{et al.} proposed a model for epidemics or opinion dynamics in \cite{HayhoeAlajajiGharesifard2018} that serves as an alternative to the famous \emph{susceptible-infected-susceptible infection model} \cite[Chapter 21]{EasleyKleinberg2010}. Specifically, they considered a contagion process on a network where every node has a private P\'{o}lya urn, but to update this urn at every time step, the node samples from a ``super urn'' which pools all the neighboring urns together. As a result, this model can account for spatial infections between neighbors. The authors then analyzed the asymptotic behavior of the proposed contagion process using martingale methods. 

The aforementioned models assume that agents interact on a fixed network, and several models have been proposed to remedy this. For example, Hanaki \emph{et al.} suggested a social network model where agents can locally alter the structure of the underlying network \cite{Hanakietal2007}, and Fazeli and Jadbabaie extended the network formation model introduced in \cite{SkyrmsPemantle2000} to develop an opinion dynamics model with a random network \cite{FazeliJadbabaie2011}. In the latter model, connections between agents are formed endogenously according to the rule ``people who have been visited are more likely to be visited in the future.'' 

In almost all of the opinion dynamics models discussed above, agents share their opinions truthfully. In contrast, agents may lie about their inherent opinions to conform with their neighbors in our model. Thus, the questions that we ask and address in this work are very different from those addressed in the literature. Whereas the classical models analyze the asymptotic behavior of opinions, we try to understand whether estimation of inherent opinions is possible in the presence of unreliable agents who are under social pressure.It is worth mentioning that in the model in \cite{ye2019influence}, agents may also express opinions that are different to their private opinions. However, our model is quite different to \cite{ye2019influence}, because our focus is on the estimation of static inherent opinions, while \cite{ye2019influence} does not consider estimation of its dynamic private opinions.

Finally, we briefly describe the classical literature on \emph{discrete choice models} from social choice theory which motivates our definition of opinion dynamics using an interacting P\'{o}lya urn model. Discrete choice models try to understand how individuals make choices between multiple alternatives, e.g., choice of their mode of transportation, choice of which opinion to broadcast, etc. One of the oldest and most widely known such models is the \emph{Bradley-Terry-Luce} (BTL) or \emph{multinomial logit} model \cite{Zermelo1929,BradleyTerry1952,Luce1959,McFadden1973}. As noted earlier, we assume that each agent's truth probability follows the BTL model \cite{Zermelo1929,BradleyTerry1952,Luce1959,McFadden1973}, where the agent's truth probability is proportional to the number of times it observes broadcast opinions that agree with its inherent opinion. Broadly speaking, the BTL model states that given $n$ choices with preference parameters $p_1,\dots,p_n \geq 0$, the probability of choosing the $i$th choice is $p_i/(p_1 + \cdots + p_n)$. The theoretical and empirical validity of this model has been demonstrated in various disciplines using various arguments. For example, Zermelo developed the rudiments of this model to accurately determine rankings of chess players \cite{Zermelo1929}; in fact, the model underlies the FIDE rating system of international chess, as well as the ranking conventions of other sports, even today. In statistics, Bradley and Terry considered the BTL model in the special case of pairwise comparisons, and Plackett formulated and studied the general BTL model from the perspective of estimating distributions over permutations \cite{Plackett1975}. In psychology, Thurstone developed a discrete choice model called the ``law of comparative judgment'' to understand how humans choose items from a set of choices \cite{Thurstone1927}, a variant of which was later shown to be equivalent to the BTL model. On the other hand, Luce defined ``Luce's choice axiom'' (or the independence axiom, as it is known in social choice theory), which essentially states that the choice of an item from a set is independent of the existence of other items in the set. He then showed that this axiom produces the BTL model \cite{Luce1959}. The microeconomics foundations of discrete choice models, such as the BTL model, were established by McFadden \cite{McFadden1973}. According to McFadden, the utility an agent obtains by choosing a certain item has deterministic and random parts. The former models the agent's and item's relevant attributes that are observable by the economist performing the study, while the latter captures the unobservable attributes. Under some assumptions on the random part, McFadden showed that the optimum choice probabilities that maximize agents' utilities are given by the BTL model \cite{McFadden1973}. The BTL model also appears in several other disciplines, e.g., maximum entropy priors in information theory, the Gibbs distribution in physics, etc. We refer readers to, e.g., \cite{JadbabaieMakurShah2020journal}, for references to many other recent works on the BTL model. Propelled by this vast literature, our definition of agents' truth probabilities in \eqref{Eq: broadcasting_probability} obeys the BTL model. This naturally begets the interacting P\'{o}lya urn structure of opinion dynamics in Section \ref{sec:model}.

\subsection{Notational Preliminaries}

We briefly introduce some useful notation for the sequel. We define $\Z_+$ and $\R_+$ to be the sets of non-negative integers and non-negative real numbers, respectively. We let $\I\{\cdot\}$ denote the indicator function, which equals $1$ if its input proposition is true, and $0$ otherwise. We denote by $\bin(n,p)$ the binomial distribution with parameters $n\in \Z_+ \backslash\{0\}$ and $p\in [0,1]$, and by $\Ber(p)$ the Bernoulli distribution with parameter $p \in [0,1]$. Lastly, we let $\P(\cdot)$ and $\E[\cdot]$ be the probability and expectation operators, where the underlying measure is clear from context.

\subsection{Outline}

The rest of the paper is organized as follows. We introduce our new opinion dynamics model and our objectives in Section \ref{sec:model}.  In Section \ref{sec:complete_phi0}, we analyze the case where the agents interact on a complete graph and have the same inherent opinions. Section \ref{sec:complete_general} studies the general complete graph setting. Section \ref{sec:conclusion} presents our conclusions and future work. Lastly, the proofs of our results are provided in the appendices.

\section{Formal Model and Goals} \label{sec:model}

We consider a set of $N$ agents denoted by $V=\{1,\dots,N\}$. Each agent $v\in V$ has an   \emph{inherent opinion}  $\phi_v \in \{0,1\}$, e.g., their support for the Democratic or Republican party. At each time step $n\in \mathbb{Z}_+$, agent $v$ broadcasts a \emph{declared opinion} $\psi_{v,n} \in \{0,1\}$ as follows: 
\begin{equation} \label{Eq: opnion}
\psi_{v,n} \triangleq 
\begin{cases}
\phi_v , & \text{with probability } p_{v,n} \\
1-\phi_v , & \text{with probability } 1-p_{v,n}.
\end{cases}
\end{equation}
According to \eqref{Eq: opnion}, agent $v$ declares its inherent opinion with probability $p_{v,n} \in (0,1)$ and lies with probability $1-p_{v,n}$. The declared opinions $\{\psi_{v,n} : v\in V, \, n \in \Z_+\}$ are broadcast on a connected undirected graph $\G = (V,E)$ (with self-loops, but no multiple edges), where the edge set $E \subseteq  V \times V$ encodes the relationship between different vertices or agents. In particular, $\{i,j\} \in E$ means that agents $i$ and $j$ can observe each other's declared opinions. For agent $v \in V$ at time $n \in \Z_+$, the truth probability (probability of broadcasting the inherent opinion) is given by the BTL model \cite{Zermelo1929,BradleyTerry1952,Luce1959,McFadden1973}:
\begin{equation} \label{Eq: broadcasting_probability}
p_{v,n} \triangleq \frac{a_{v,n}}{a_{v,n} + b_{v,n}} \,,
\end{equation}
where the random parameters $a_{v,n},b_{v,n} \in \R_+$ are updated as follows:
\begin{enumerate}
\item At time $n=0$, $a_{v,0} = b_{v,0} = 1$ for all $v \in V$.
\item At time $n \geq 1$, every agent $v \in V$ observes its neighbors' declared opinions $\{\psi_{u,n-1} : u \in \N_v\}$, where $\N_v = \{u \in V : \{v,u\} \in E\}$ denotes its set of neighbors. Then, it updates its truth probability parameters as an interacting P\'{o}lya urn process (with reinforcement) \cite[Section 2.4]{LevinPeresWilmer2009}, \cite{Pemantle2007}:
\begin{align}
a_{v,n} & = a_{v,n-1} + \gamma \underbrace{\sum_{u \in \N_v}{\I\!\left \{\psi_{u,n-1}=\phi_v\right\}}}_{\text{declared opinions that equal } \phi_v} , \label{Eq: parameter_a} \\
b_{v,n} & = b_{v,n-1} + \sum_{u \in \N_v}{\I\!\left\{\psi_{u,n-1}\neq \phi_v\right\}} ,\label{Eq: parameter_b}
\end{align}
where $\gamma > 0$ is a known \emph{honesty} parameter.
\end{enumerate}
Note that $a_{v,0}$ and $b_{v,0}$ can take any strictly positive values without changing our analysis. We set them to be $1$ to simplify the presentation of our results.

According to \eqref{Eq: broadcasting_probability}, \eqref{Eq: parameter_a}, and \eqref{Eq: parameter_b}, at every time step, agent $v$ must either declare $\phi_v$ or lie about it. It makes this decision by conforming to its neighbors'  declared opinions. In fact, unwinding \eqref{Eq: parameter_a} and \eqref{Eq: parameter_b} and noting that the dominant terms in $a_{v,n}$ and $b_{v,n}$ are $\gamma \sum_{t=0}^n \sum_{u \in \N_v}{\I\!\left \{\psi_{u,t-1}=\phi_v\right\}}$ and $\sum_{t=0}^n \sum_{u \in \N_v}{\I\!\left \{\psi_{u,t-1}\neq \phi_v\right\}}$ (see \propref{Prop: Non-existence of Almost Sure Limit}), we obtain that 
\begin{align*}
    &\frac{\text{ probability of telling the truth at time $n$}}{\text{probability of lying at time $n$}} \approx \\
    & \qquad \qquad \gamma \times  \frac{\#\!\left\{\text{\parbox{5cm}{ declared opinions that confirm $v$'s inherent opinion until time $n$}}\right\}}{\#\!\left\{\text{\parbox{5cm}{ declared opinions that contradict $v$'s inherent opinion until time $n$}}\right\}}.
\end{align*}
Hence, the more the agent observes a certain opinion, the more it is forced to declare this opinion. To understand the meaning of the honesty parameter $\gamma$, let us consider a scenario where the social pressure on an agent is neutral, i.e., it observes an equal number of $0$ and $1$ declared opinions. In this case, the ratio of the probability of telling the truth to the probability of lying is $\gamma$. This means that the higher the value of $\gamma$, the more honest the agent. In particular, if $\gamma > 1$ (respectively, $\gamma < 1$), the agent leans towards telling the truth (respectively, lying). We will assume throughout this paper that $\gamma >1$. We remark that such honesty (or conformity) parameters have been considered in the literature in the setting of Friedkin and Johnsen's model, cf. the work of \cite{das2014role} in this setting.

The aforementioned description completely determines a stochastic process with reinforcement on $\G$ along with its initial conditions. This raises the natural question as to whether this process is ergodic, or more generally, whether any information about $\{\phi_v : v \in V\}$ can be inferred from observations of the declared opinions $\H_{n} \triangleq \{\psi_{v,k} : v \in V, \, 0 \leq k < n\}$ after an arbitrarily long period of time, i.e., as $n \rightarrow \infty$. In other words, can one infer the agents' inherent opinions despite the fact that the agents may lie to conform with their neighbors? 

Under possible regularity conditions, our objectives are to:
\begin{enumerate}
\item Estimate the \emph{proportion of agents} that have inherent opinions equal to $1$, viz.
\begin{equation} 
\label{Eq: fraction of 1}
\Phi \triangleq \frac{1}{N}\sum_{u \in V}{\phi_u} 
\end{equation}
based on $\H_{n}$ as $n \rightarrow \infty$. Formally, we seek to construct strongly consistent estimators $f_n:\{0,1\}^{N \times n} \rightarrow [0,1]$ such that
\begin{equation}
\label{Eq: goal 1}
\P\!\left(\lim_{n \rightarrow \infty}{f_n(\H_n)} = \Phi\right) = 1 , 
\end{equation}
where underlying probability law is defined by the random variables $\{\psi_{v,k} : v \in V, \, k \in \Z_+\}$. In the earlier example where agents' inherent opinions are their true political alliances and their declared opinions are political inclinations of tweets or posts on  social media, the problem of learning $\Phi$ corresponds to predicting election results by observing voters' declared opinions.
\item Decode all inherent opinions $\{\phi_v : v \in V\}$ from $\H_{n}$ as $n \rightarrow \infty$. Formally, we also seek to construct strongly consistent estimators $h_n : \{0,1\}^{N \times n} \rightarrow [0,1]^{N}$ such that
\begin{equation}
\P\!\left(\lim_{n \rightarrow \infty}{h_n(\H_n)} = \{\phi_v : v \in V\}\right) = 1 .
\end{equation}
\end{enumerate}

While the aforementioned interacting P\'{o}lya urn model is one of the first models that captures the important phenomenon of agents possessing two kinds of opinions (to the best of our knowledge), it has several limitations that we briefly delineate next. Firstly, we assume that inherent opinions are static (or constant over time). This is only an accurate assumption in situations where all individuals are very dogmatic, and models where inherent opinions change over time have also been proposed in the literature, cf. \cite{ye2019influence}. Secondly, we assume that the honesty parameter $\gamma$ is the same for all agents, and is \emph{known to the estimator}. Thirdly, we assume that the manner in which agents influence each other is symmetric and homogeneous, e.g., the underlying graph is undirected. This need not be the case in real-world settings. Furthermore, as noted in \secref{Contributions}, we only analyze the complete graph case in Sections \ref{sec:complete_phi0} and \ref{sec:complete_general}. Although somewhat unrealistic, these assumptions make the problem analytically tractable and allow us to draw conclusions on whether it is possible to estimate inherent opinions.

\section{Analysis of Complete Graph and a Homogeneous Population} \label{sec:complete_phi0}

In this section, we analyze the special case where the graph $\G$ is complete with self-loops, and the agents are homogeneous in the sense that they have equal inherent opinions. Moreover, we assume that the estimator is aware of the population's homogeneity. So, the estimator's goal is to decide whether all agents have inherent opinions equal to $0$ or $1$. Although this is a contrived setting, it affords us several insights about the choice of estimator and the technical tools used to prove consistency that will be very useful in the more general setting.

Without loss of generality, we assume that all the agents have inherent opinion $0$, i.e., $\phi_v = 0$ for all $v \in V$. In this case, the dynamics of $a_{v,n}$ and $b_{v,n}$ are independent of the agent $v$. Hence, dropping the subscript $v$ from $a_{v,n}$, $b_{v,n}$, and $p_{v,n}$, we may write
\begin{equation}
\label{Eq: Initial Conditions}
a_0 = b_0 = 1, 
\end{equation}
so that $p_0 = \frac{1}{2}$, and for all $n \geq 1$, $p_n = a_n/(a_n + b_n)$ with
\begin{align}
\label{Eq: Dynamics of a}
a_{n} & = a_{n-1} + \gamma N \left(1 - \Psi_{n-1}\right), \\
b_{n} & = b_{n-1} + N \Psi_{n-1},
\label{Eq: Dynamics of b}
\end{align}
where we define the random variables:
\begin{equation} \label{Eq: average opinion}
\forall n \in \Z_+, \enspace \Psi_n \triangleq \frac{1}{N}\sum_{u \in V}{\psi_{u,n}} \in \left\{0,\frac{1}{N},\frac{2}{N},\dots,1\right\} . 
\end{equation}
$\Psi_n$ is the fraction of agents that hold a declared opinion of $1$ at time $n$.

Consider the sequence of estimators:
\begin{equation}
\label{Eq: Estimator Definition}
\forall n \geq 1, \enspace f_n(\H_n) = \hat{\Psi}_n \triangleq \frac{1}{n} \sum_{k=0}^{n-1}{\Psi_k} \, .
\end{equation}
To prove \eqref{Eq: goal 1}, it suffices to establish that $\hat{\Psi}_n \rightarrow 0$ as $n \rightarrow \infty$ almost surely ($a.s.$). Hence, we now analyze the almost sure convergence of the sequence of random variables $\{\hat{\Psi}_n : n \geq 1\}$. This is done in three steps. First, we show in \propref{Prop: Martingale Property} that the truth probability $p_n$ converges $a.s.$ to some random variable $p^*$. Second, in \thmref{Thm: Convergence to Bernoulli Variable}, we establish that the limit
$p^*$ belongs to $\{0,1\}$, i.e., $p^*$ is  Bernoulli. Moreover, we prove that the average opinion $\hat \Psi_n$ converges $a.s.$ to $1-p^*$. Finally, \thmref{Thm: Asymptotic Learning} conveys that $p^*$ is indeed equal to $1$ $a.s.$, which means that the agents tell the truth in the limit. 

For any $n \in \Z_+$, conditioned on $a_n$ and $b_n$, $N \Psi_n \sim \bin(N,1-p_n)$ obeys a binomial distribution:
\begin{equation}
\label{Eq: Transition of Psi_t}
 \P\!\left(\Psi_{n} = \frac{j}{N} \, \middle| \, a_n,b_n\right) = \binom{N}{j} \left(\frac{b_n}{a_n + b_n}\right)^{j} \left(\frac{a_n}{a_n + b_n}\right)^{N-j} 
\end{equation}
for all $j \in \{0,\dots,N\}$, because $\{\psi_{n,u} : u \in V\}$ are conditionally independent and identically distributed $\Ber(1 - p_n)$ random variables. Together, the initial conditions in \eqref{Eq: Initial Conditions} and the dynamics defined by \eqref{Eq: Dynamics of a}, \eqref{Eq: Dynamics of b}, and \eqref{Eq: Transition of Psi_t} characterize the trajectory of $\{(a_n,b_n,\Psi_n) : n \in \Z_+\}$. Observe that the truth probability $p_n$ is invariant to scaling of $(a_n,b_n)$. So, consider the normalized probability parameters $\alpha_n \triangleq a_n / N$ and $\beta_n \triangleq b_n / N$ for all $n \in \Z_+$, so that $p_n = \alpha_n/(\alpha_n + \beta_n)$. These normalized parameters satisfy the stochastic dynamics:
\begin{align}
\label{Eq: Dynamics of alpha}
\alpha_{n} & = \alpha_{n-1} + \gamma \left(1 - \Psi_{n-1}\right) , \\
\beta_{n} & = \beta_{n-1} + \Psi_{n-1} \, ,
\label{Eq: Dynamics of beta} \\ \P\!\left(\Psi_{n} = \frac{j}{N} \, \middle| \, \alpha_n,\beta_n\right) & = \binom{N}{j} (1-p_n)^{j} p_n^{N-j} ,
\label{Eq: Modified Transition of Psi_t}
\end{align}
for all $j \in \{0,\dots,N\}$, with initial conditions $\alpha_0 = \beta_0 = \frac{1}{N}$. It should be noted that $(\alpha_n,\beta_n)$ forms a Markov process.

In the remainder of this
section, we show that the 
sequence of estimators 
$f_n$ learns that all agents 
have inherent opinion $0$ over time. Let $\F_n \triangleq \sigma(\H_n)$ be the smallest $\sigma$-algebra generated by $\H_n$, and $\F=\{\F_n : n \in \Z_+\}$ be the corresponding filtration. The ensuing proposition constructs martingales and shows that $p_n$ converges $a.s.$ as $n \rightarrow \infty$.

\begin{Proposition}[Martingale Property]
\label{Prop: Martingale Property}
The following hold:
\begin{enumerate}
\item $\{p_n : n \in \Z_+\}$ is an $\F$-sub-martingale, or equivalently, $\{1-p_n : n \in \Z_+\}$ is an $\F$-super-martingale.
\item There exists a random variable $p^* \in [0,1]$ such that
$$ p^* = \lim_{n \rightarrow \infty}{p_n} \enspace a.s. $$
\end{enumerate}
\end{Proposition}

\propref{Prop: Martingale Property} is established in \appref{Proof of Proposition Martingale Property}. While one might intuitively expect something like \propref{Prop: Martingale Property} to hold due to the close resemblance of $\{p_n : n \in \Z_+\}$ with a \emph{P\'{o}lya urn process} (which is obtained by setting $\gamma = 1$ and $N= 1$), we note that unlike the P\'{o}lya urn process, $\{\Psi_n : n \in \Z_+\}$ is not an exchangeable sequence (cf. \cite[Sections VII.4 and VII.9]{Feller1971}, \cite[Sections 2.2 and 2.4]{Pemantle2007}). Hence, one cannot apply de Finetti's theorem to easily deduce the probability law of the limiting random variable. Furthermore, we remark that since $p_0 = \frac{1}{2}$ $a.s.$ and the sub-martingale property in \propref{Prop: Martingale Property} implies that $\{\E[p_n] : n \in \Z_+\}$ forms a non-decreasing sequence, we obtain that
\begin{equation}
\label{Eq: Expected Value Bounds}
\lim_{n \rightarrow \infty}{\E[p_n]} = \E[p^*] \in \left[\frac{1}{2},1\right]
\end{equation}
using Lebesgue's dominated convergence theorem \cite[Chapter 1, Theorem 4.16]{Cinlar2011}. We next demonstrate using \propref{Prop: Martingale Property} that the limiting random variable $p^*$ in part 2 of \propref{Prop: Martingale Property} is actually a Bernoulli random variable, and it characterizes the limits of both $\hat{\Psi}_n$ and $\Psi_n$ in appropriate senses.

\begin{Theorem}[Convergence to Bernoulli Variable]
\label{Thm: Convergence to Bernoulli Variable}
The following are true:
\begin{enumerate}
\item $p^* \in \{0,1\}$ $a.s.$
\item $\lim_{n \rightarrow \infty}{\hat{\Psi}_n} = 1 - p^*$ $a.s.$
\item $\Psi_n \stackrel{w}{\rightarrow} 1-p^*$ as $n \rightarrow \infty$, i.e., $\Psi_n$ converges weakly (or in distribution) to $1-p^*$.
\end{enumerate}
\end{Theorem}

\thmref{Thm: Convergence to Bernoulli Variable} is proved in \appref{Proof of Theorem Convergence to Bernoulli Variable}. Upon inspecting part 3, it is natural to wonder whether the simpler estimator $\Psi_n$ learns the true inherent opinions. In the ensuing proposition, we show that the answer is negative; in particular, $\Psi_n$ does not converge almost surely to $1-p^*$ although part 3 of \thmref{Thm: Convergence to Bernoulli Variable} establishes the weak convergence.

\begin{Proposition}[Non-existence of Almost Sure Limit]
\label{Prop: Non-existence of Almost Sure Limit}
$\{\Psi_n : n \in \Z_+\}$ does not converge to $1-p^*$ $a.s.$ as $n \rightarrow \infty$. Specifically, we have
$$ \P\!\left(\lim_{n \rightarrow \infty}{\Psi_n} \in \{0,1\}\right) = 0 \, , $$
i.e., the probability that $\lim_{n \rightarrow \infty}{\Psi_n}$ exists and belongs to $\{0,1\}$ is zero.
\end{Proposition}

The proof of \propref{Prop: Non-existence of Almost Sure Limit} can be found in \appref{Proof of Proposition Non-existence of Almost Sure Limit}. Finally, in the next theorem, we establish that $p^* = 1$ $a.s.$

\begin{Theorem}[Asymptotic Learning I]
\label{Thm: Asymptotic Learning}
The following hold:
\begin{enumerate}
\item $p^* = 1$ $a.s.$
\item $\lim_{n \rightarrow \infty}{\hat{\Psi}_n} = 0$ $a.s.$
\end{enumerate}
\end{Theorem}

Theorem \ref{Thm: Asymptotic Learning} is derived using \emph{stochastic approximations} in \appref{Proof of Theorem Asymptotic Learning}. It conveys that if our estimator $f_n$ is aware that the population is homogeneous, then it recovers $\Phi$ in the sense of \eqref{Eq: goal 1}. In particular, when $\hat \Psi_n$ converges to $0$ (respectively $1$), all agents have inherent opinions equal to $0$ (respectively $1$). Intuitively, our results in Proposition \ref{Prop: Non-existence of Almost Sure Limit} and Theorem \ref{Thm: Asymptotic Learning} show that making accurate predictions regarding which inherent opinion is held by the majority based on a poll at some particular instance in time $n$ is not possible. Indeed, it could be that although $\Phi \geq \frac{1}{2}$, there is an arbitrarily large instance in time $n$ where $\Psi_n < \frac{1}{2}$. However, over time, our estimator $\hat{\Psi}_n$ would correctly show that $1$ is the majority inherent opinion.

\section{Analysis of Complete Graph with  $\Phi \in [0,1]$} \label{sec:complete_general}

In this section, we analyze the complete graph (with self-loops) case for general $\Phi$. In this setting, the dynamics of $a_{v,n}$ and $b_{v,n}$ depend only on the inherent opinion $\phi_v$ of agent $v$. In particular, agents $v$ with $\phi_v = j \in \{0,1\}$ have common probability parameters $a_n^j$ and $b_n^j$ that evolve as follows: 
\begin{align}
    a_n^0 & = a_{n-1}^0+ \gamma N (1-\Psi_{n-1}) , \label{Eq: Dynamics of a0} \\
    a_n^1 & = a_{n-1}^1 + \gamma N \Psi_{n-1} ,  \label{Eq: Dynamics of a1} \\ 
    b_n^0 & = b_{n-1}^0 + N \Psi_{n-1} , \label{Eq: Dynamics of b0} \\
    b_n^1 & = b_{n-1}^1 + N  (1-\Psi_{n-1}) , \label{Eq: Dynamics of b1} 
\end{align}
for all $n \geq 1$, with $a_0^0= a_0^1 = b_0^0 = b_0^1 = 1$. Moreover, agents with inherent opinion $j\in\{0,1\}$ tell the truth at time $n \in \Z_+$ with probability 
\begin{equation} \label{Eq: probability pj}
    p_{n}^j=\frac{a_{n}^j}{a_n^j+b_n^j}.
\end{equation}
We define the normalized probability parameters $\alpha_n^j\triangleq a_n^j / N$ and $\beta_n^j\triangleq b_n^j / N$ for all $n\in \Z_+$ and $j \in \{0,1\}$. Thus, $p_n^j=\alpha_n^j/(\alpha_n^j+\beta_n^j)$ as before. These normalized parameters satisfy the following dynamics:
\begin{align}
    \alpha_n^0 & = \alpha_{n-1}^0+ \gamma  (1-\Psi_{n-1}) , \label{Eq: Dynamics of alpha0}\\
    \alpha_n^1 & = \alpha_{n-1}^1 + \gamma \Psi_{n-1} ,  \label{Eq: Dynamics of alpha1}\\ 
    \beta_n^0 & = \beta_{n-1}^0 +  \Psi_{n-1} , \label{Eq: Dynamics of beta0}\\
    \beta_n^1 & = \beta_{n-1}^1 +  (1-\Psi_{n-1}). \label{Eq: Dynamics of beta1}
\end{align}
As before, note that $(\alpha_n^0,\alpha_n^1,\beta_n^0,\beta_n^1)$ forms a Markov process. We denote by $V^j$ the set of agents that have inherent opinion $j \in \{0,1\}$, and define $\Psi_n^j \triangleq \sum_{v \in V^j} \psi_{v,n} / |V^j|$ for all $n \in \Z_+$ and $j\in \{0,1\}$. Thus, we may write: 
\begin{align}
    \Psi_n & = (1-\Phi) \Psi_n^0 + \Phi \Psi_n^1 \, , \label{Eq: psi_n}\\
  \P\!\left(\Psi_{n}^0 = \frac{i}{|V^0|} \, \middle| \, \alpha_n^0,\beta_n^0\right) & = \binom{|V^0|}{i} \left(1-p_n^0\right)^{i} \left(p_n^0\right)^{|V^0|-i} , \\
     \P\!\left(\Psi_{n}^1 = \frac{j}{|V^1|} \, \middle| \, \alpha_n^1,\beta_n^1\right) & = \binom{|V^1|}{j} \left(p_n^1\right)^{j} \left(1-p_n^1\right)^{|V^1|-j} , \\
    \E\!\left[\Psi_n \, \middle| \, \alpha_n^0,\alpha_n^1,\beta_n^0,\beta_n^1  \right] &= (1-\Phi) (1-p_n^0) + \Phi p_n^1 \, , \label{Eq: Epsi_n} 
\end{align}
for all $i \in \{0,\dots,|V^0|\}$ and $j \in \{0,\dots,|V^1|\}$, where $\Phi = |V^1|/N$ is the fraction of agents that have inherent opinion $1$. 

We can construe \eqref{Eq: Dynamics of a0}-\eqref{Eq: Dynamics of b1} as the equations describing the dynamics of two interacting P\'{o}lya urns $u_0$ and $u_1$. At time $n$, $u_0$ contains $a^0_n$ white balls and $b^0_n$ red balls, while $u_1$ contains $a^1_n$ white balls and $b^1_n$ red balls. At the next time step $n+1$, $|V^j|$ balls are chosen randomly (with replacement) from $u_j$ for $j\in \{0,1\}$, and we let $W_n = N (1 - \Psi_n)$ and $R_n = N \Psi_n$ be the total number of white and red balls chosen, respectively. Subsequently, $\gamma W_n$ white balls are added to $u_0$, $\gamma R_n$ white balls are added to $u_1$, $R_n$ red balls are added to $u_0$, and $W_n$ red balls are added to $u_1$. Analyzing the asymptotic behavior of two such interacting urns using stochastic approximations is challenging. Indeed, the probabilities $p^0_n$ and $p_n^1$ of choosing a white ball satisfy the following equations for all $n \in \Z_+$:
\begin{align} 
    p_{n+1}^0- p_n^0 & = \frac{1}{\alpha_{n}^0+\beta_n^0+\gamma (1-\Psi_n)+\Psi_n} \! \left( F^0(p_n^0,p_n^1) + \xi_n^0 \right)\label{Eq: evolution p0}\\
    p_{n+1}^1-p_n^1 & = \frac{1}{\alpha_n^1+\beta_n^1 + \gamma \Psi_n + 1-\Psi_n} \left ( F^1(p_n^0,p_n^1) + \xi_n^1 \right )\label{Eq: evolution p1}
\end{align}
where
\begin{align*}
F^0(x,y) &= -(\gamma-1) (1-\Phi) x^2 + (\gamma -1 )\Phi xy  \\ &+\left (   (\gamma -1 )(1-\Phi) - \gamma \Phi  \right)x- \gamma \Phi y + \gamma \Phi , \\ 
\xi_n^0 &= \gamma(1-\Psi_n) - p_n^0 (\gamma (1-\Psi_n)+\Psi_n) -F^0(p_n^0,p_n^1) , \\
F^1(x,y)&= -(\gamma -1 )\Phi y^2 +(\gamma -1) (1-\Phi)  xy \, + 
\\& \left (   (\gamma -1 )\Phi- \gamma (1-\Phi)\right) y - \gamma (1-\Phi) x + \gamma (1-\Phi) , \\
\xi_n^1 &= \gamma \Psi_n - p_n^1 (\gamma \Psi_n + 1-\Psi_n)  - F^1(p_n^0,p_n^1),
\end{align*}
for $x,y \in [0,1]$. The step sizes $ (\alpha_{n}^0+\beta_n^0+\gamma (1-\Psi_n)+\Psi_n)^{-1}$ and $(\alpha_n^1+\beta_n^1 + \gamma \Psi_n + 1-\Psi_n)^{-1}$ are random and different, which makes the stochastic approximation of \eqref{Eq: evolution p0}-\eqref{Eq: evolution p1} by a differential equation rather difficult. 

So, instead of considering \eqref{Eq: Dynamics of a0}-\eqref{Eq: Dynamics of b1} as the dynamics of two urns with two colors, we perceive them as describing one urn with four colors (white, red, blue, and yellow). Following this interpretation, the model becomes tractable. Indeed, the number of white, red, blue, and yellow balls at time $n$ are $a_n^0$, $b_n^0$, $a_n^1$, and $b_n^1$, respectively. Define the probabilities of drawing a white, red, or blue ball as, respectively: 
\begin{align}
    X_n & \triangleq \frac{\alpha_n^0}{\alpha_n^0+\beta_n^0+\alpha_n^1+\beta_n^1} , \label{Eq: definition X} \\
    Y_n & \triangleq \frac{\beta_n^0}{\alpha_n^0+\beta_n^0+\alpha_n^1+\beta_n^1} , \label{Eq: definition Y}\\
    Z_n & \triangleq \frac{\alpha_n^1}{\alpha_n^0+\beta_n^0+\alpha_n^1+\beta_n^1} . \label{Eq: definition Z} 
\end{align}
The process $(X_n,Y_n,Z_n)$ satisfies the following equations:
\begin{align}
    &X_{n+1} - X_n  \nonumber\\ &=  \frac{\alpha_n^0+ \gamma (1-\Psi_n)}{\alpha_n^0+\beta_n^0+\alpha_n^1+\beta_n^1+\gamma+1} -  \frac{\alpha_n^0}{\alpha_n^0+\beta_n^0+\alpha_n^1+\beta_n^1} \nonumber\\
    & =\frac{1}{\alpha_n^0+\beta_n^0+\alpha_n^1+\beta_n^1+\gamma+1} \left(  \gamma (1-\Psi_n) -  (\gamma+1)X_n\right) \nonumber\\
    & = \gamma_{n+1} \left( G^1(X_n,Y_n,Z_n)+U_{n+1}^1\right), \label{Eq: SA-1}\\
    &Y_{n+1}-Y_n  \nonumber \\ &= \frac{\beta_n^0+ \Psi_n}{\alpha_n^0+\beta_n^0+\alpha_n^1+\beta_n^1+\gamma+1} -  \frac{\beta_n^0}{\alpha_n^0+\beta_n^0+\alpha_n^1+\beta_n^1} \nonumber\\
    & =\frac{1}{\alpha_n^0+\beta_n^0+\alpha_n^1+\beta_n^1+\gamma+1} \left( \Psi_n -  (\gamma+1)Y_n\right) \nonumber\\
    & = \gamma_{n+1} \left( G^2(X_n,Y_n,Z_n)+U_{n+1}^2\right), \label{Eq: SA-2}\\
    &Z_{n+1}-Z_n \nonumber   \\
    &= \frac{\alpha_n^1+\gamma \Psi_n}{\alpha_n^0+\beta_n^0+\alpha_n^1+\beta_n^1+\gamma+1} - \frac{\alpha_n^1}{\alpha_n^0+\beta_n^0+\alpha_n^1+\beta_n^1}\nonumber\\
    & = \frac{1}{\alpha_n^0+\beta_n^0+\alpha_n^1+\beta_n^1+\gamma+1}\left( \gamma \Psi_n - (\gamma+1)Z_n\right)  \nonumber\\
    & = \gamma_{n+1} \left( G^3(X_n,Y_n,Z_n) +U^3_{n+1} \right) , \label{Eq: SA-3}
\end{align}
for all $n \in \Z_+$, where 
\begin{align}
\gamma_{n+1} & =
\frac{1}{(4/N) +n(\gamma+1)} ,  \label{Eq: gamma_n}\\
    G^1(x,y,z) &= -(\gamma+1)x \, + \nonumber \\
    & \gamma \!\left(\! (1-\Phi) \frac{x}{x+y} + \Phi \left( 1- \frac{z}{1-x-y} \right) \!\right) \! , \label{Eq: G1} \\
    U^1_{n+1}&=\gamma (1-\Psi_n) -  (\gamma+1)X_n - G^1(X_n,Y_n,Z_n) , \label{Eq: U1}\\
    G^2(x,y,z)& = -(\gamma+1)y \, + \nonumber \\ & (1-\Phi) \left(1-\frac{x}{x+y} \right) + \Phi \frac{z}{1-x-y} , \label{Eq: G2} \\
    U^2_{n+1} &= \Psi_n -  (\gamma+1)Y_n - G^2(X_n,Y_n,Z_n) , \label{Eq: U2}\\
    G^3(x,y,z) & = -(\gamma+1) z \, + \nonumber \\ & \gamma \!\left(\! (1-\Phi) \left( 1- \frac{x}{x+y} \right) + \Phi \frac{z}{1-x-y}\!\right) \! , \label{Eq: G3} \\
    U^3_{n+1} & = \gamma \Psi_n - (\gamma+1)Z_n - G^3(X_n,Y_n,Z_n). \label{Eq: U3}
\end{align}
Equations \eqref{Eq: SA-1}-\eqref{Eq: SA-3} have a common step size $\gamma_{n+1}$, which allows us to approximate \eqref{Eq: SA-1}-\eqref{Eq: SA-3} using the following \emph{ordinary differential equation} (ODE):
\begin{align} \label{Eq: ODE}
 \frac{d}{dt}\begin{pmatrix}
   x \\ y \\z
   \end{pmatrix} =  G(x,y,z) \triangleq
   \begin{pmatrix}
    G^1(x,y,z) \\ G^2(x,y,z) \\  G^3(x,y,z)
   \end{pmatrix}.   
\end{align}
It should be noted that $p_n^0=X_n/(X_n+Y_n)$ and $p_n^1=Z_n/(1-X_n-Y_n)$. This means that analyzing the stochastic process $(X_n,Y_n,Z_n)$ allows us to deduce the asymptotic behavior of the process of interest $(p_n^0,p_n^1)$.

At first glance, the ODE \eqref{Eq: ODE} seems to have singularities at $x+y=0$ and $x+y=1$. However, since  \eqref{Eq: ODE} is introduced to approximate the process $(X_n,Y_n,Z_n)$, it is sufficient to show that $G$ is smooth on the range of $(X_n,Y_n,Z_n)$, and that any solution of \eqref{Eq: ODE} that starts in this range remains in the range, i.e., the range of $(X_n,Y_n,Z_n)$ is an invariant set of the ODE. These facts are derived in the following lemma.

\begin{Definition}[Invariant Set]
 A set $M \subseteq \R^m$ is an \emph{invariant} set of the ODE $dx/dt = f(x)$ with vector field $f : \R^n \rightarrow \R^n$ if for all initial conditions $x(0)\in M$ and all $t \in \R_+$, the solution $x(t)$ of the ODE lives in $M$, i.e., $x(t)\in M$.
\end{Definition}

\begin{Lemma}[Well-defined ODE] \label{lem: domains}
The following are true:
\begin{enumerate}
    \item The stochastic processes $X_n$, $Y_n$, $Z_n$, and $X_n+Y_n$ belong to $\big[0,\frac{\gamma}{\gamma+1}\big]$, $\big[0,\max\!\big(\frac{1}{\gamma+1},\frac{1}{4}\big)\big]$, $\big[0,\frac{\gamma}{\gamma+1}\big]$, and $\big[\frac{1}{\gamma+1},\frac{\gamma}{\gamma+1}\big]$, respectively.
    \item The compact manifold
    \begin{align*}
        &M \triangleq \Bigg \{(x,y,z)\in \R^3 : x \in \left [0,\frac{\gamma}{\gamma+1}\right ]\!, \, z \in \left [0,\frac{\gamma}{\gamma+1}\right ]\!, \\&y\in \left [0,\max\left (\frac{1}{\gamma+1},\frac{1}{4}\right )\right ]\!, \, x+y \in  \left [\frac{1}{\gamma+1},\frac{\gamma}{\gamma+1}\right ]\Bigg \}
    \end{align*}
    is an invariant set of the ODE \eqref{Eq: ODE}.
    \item The vector field $G$ is complete on $M$, i.e., the ODE \eqref{Eq: ODE} with initial condition in $M$ admits a unique solution on $\R_+$.
\end{enumerate}
\end{Lemma}

\lemref{lem: domains} is proved in \appref{Proof of Lemma domains}. In the remainder of this section, we show that the average declared opinion \eqref{Eq: Estimator Definition} can be used to construct an estimator that learns $\Phi$ provided that the population 
does not include large majorities, i.e., $\Phi$ is bounded away from $0$ and $1$. This main result will be given in Theorem \ref{Thm: global estimation}. Furthermore, Corollary \ref{Cor: individual estimation} will demonstrate that an estimate of $\Phi$ can be used to estimate the individual inherent opinions of the agents. To this end, in the next lemma, we start by showing that the process $(X_n,Y_n,Z_n)$ converges to the \emph{equilibrium points} of the ODE \eqref{Eq: ODE}.

\begin{Lemma}[Convergence to Equilibria] \label{lem: equilrium points}
 The ensuing statements are true:
 \begin{enumerate}
     \item The set of equilibria $L$ of \eqref{Eq: ODE} is 
     $$L=\begin{cases}
      \left\{ l_1, l_2, l_3 \right\}, & \text{if } \gamma \geq  \max\!\left( \frac{\Phi}{1-\Phi},\frac{1-\Phi}{\Phi}\right)\\
      \left\{ l_1, l_2 \right\}, & \text{otherwise} 
     \end{cases}$$
    where $l_1 =\big(0,\frac{1}{\gamma+1},\frac{\gamma}{\gamma+1}\big) $, $l_2 = \big(\frac{\gamma}{\gamma+1},0,0\big)$, and 
    $l_3 = \big(\frac{\gamma(\gamma(1-\Phi)-\Phi)}{\gamma^2-1},\frac{( \gamma \Phi - (1-\Phi))}{\gamma^2-1},\frac{\gamma ( \gamma \Phi - (1-\Phi) )}{\gamma^2-1}\big)$.
         \item The process $(p_n^0,p_n^1)$ converges $a.s.$ to $(p_\infty^0,p_\infty^1)\in L_p$ as $n \rightarrow \infty$, where 
         $$ L_p = \begin{cases}
         \left\{ (0,1),(1,0),l_3^{\prime}\right\}, & \text{if } \gamma \geq \max\!\left( \frac{\Phi}{1-\Phi},\frac{1-\Phi}{\Phi}\right) \\
         \left\{ (0,1),(1,0) \right\}, & \text{otherwise}
         \end{cases} $$ 
         and $l_3^{\prime} = \big( \frac{\gamma(\gamma(1-\Phi)-\Phi)}{(1-\Phi)(\gamma^2-1)},\frac{\gamma(\gamma\Phi-(1-\Phi))}{\Phi(\gamma^2-1)} \big)$.
 \end{enumerate}
\end{Lemma}

\lemref{lem: equilrium points} is established in \appref{Proof of Lemma equilrium points}. It conveys that the process $(p_n^0,p_n^1)$ converges to a random variable supported on a finite set. The next theorem shows that this process in fact converges to a deterministic vector, which will be used to estimate $\Phi$.

\begin{Theorem}[Convergence to Deterministic Vector] \label{Thm: convergence}
The process $(p_n^0,p_n^1)$ converges $a.s.$ to $(p_\infty^0,p_\infty^1)$, where
\begin{equation} \label{Eq: p_infty}
  (p_\infty^0,p_\infty^1) = 
  \begin{cases}
   l_3^{\prime} ,  & \text{if } \gamma \geq  \max\!\left( \frac{\Phi}{1-\Phi},\frac{1-\Phi}{\Phi}\right)\\
      (1,0) , & \text{if }  \frac{\Phi}{1-\Phi}<\gamma < \frac{1-\Phi}{\Phi}\\
      (0,1) , & \text{if }  \frac{1-\Phi}{\Phi}<\gamma < \frac{\Phi}{1-\Phi}
  \end{cases}   
\end{equation}
and $l_3^{\prime}$ is defined in \lemref{lem: equilrium points}.
\end{Theorem}

\thmref{Thm: convergence} is derived in \appref{Proof of Theorem convergence}. So far, we have illustrated that the process $(p_n^0,p_n^1)$ converges to a deterministic vector $(p_\infty^0,p_\infty^1)$. As mentioned earlier, our estimator will use the quantity \eqref{Eq: Estimator Definition} to learn $\Phi$. So, the ensuing lemma relates the asymptotic behavior of \eqref{Eq: Estimator Definition} to $(p_\infty^0,p_\infty^1)$. It will also be used to infer the individual agents' inherent opinions $\{\phi_v : v\in V\}$.

\begin{Lemma}[Individual Opinions I] \label{lemma: individual convergence}
The following are true:
\begin{enumerate}
    \item For any agent $v \in V$, its declared opinions satisfy
    \begin{equation} \label{Eq: partition}
\psi_{v,\infty} \triangleq \lim\limits_{n \to \infty} \frac{1}{n} \sum_{k=0}^{n-1}\psi_{v,k} =
\begin{cases} 
p_{\infty}^1 \, , & \phi_v = 1 \, , \\
1 - p_{\infty}^0 \, , & \phi_v = 0 \, , 
\end{cases}
\end{equation} \\
where $(p_\infty^0,p_\infty^1)$ is given by \eqref{Eq: p_infty}.
\item The average declared opinion \eqref{Eq: Estimator Definition} converges $a.s.$ to
\begin{equation} \label{Eq: lim estimator}
 \hat{\Psi}_\infty \triangleq   \lim\limits_{n \to \infty}\hat{\Psi}_n = (1-\Phi) (1-p_\infty^0) + \Phi p_\infty^1.
\end{equation}
\end{enumerate}
\end{Lemma}

The proof of \lemref{lemma: individual convergence} can be found in \appref{Proof of Lemma individual convergence}. Now consider the sequence of estimators:
\begin{equation}
\label{Eq: Estimator_general}
\forall n \geq 1, \enspace g_n(\H_n) = \frac{\hat{\Psi}_n(\gamma -1)+1}{\gamma+1}, 
\end{equation}
where $\hat{\Psi}_n$ is defined in \eqref{Eq: Estimator Definition}. We finally state our main result. 

\begin{Theorem}[Asymptotic Learning II] \label{Thm: global estimation}
The following hold:
\begin{enumerate}
\item $g_n(\H_n)$ converges $a.s.$ to a deterministic quantity $g_\infty \in \big[\frac{1}{\gamma+1},\frac{\gamma}{\gamma+1}\big]$ as $n \rightarrow \infty$. 
    \item If $g_\infty = \frac{1}{\gamma+1}$, then $\Phi \leq \frac{1}{\gamma+1}.$
    \item If $g_\infty =\frac{\gamma }{\gamma+1}$, then 
    $\Phi \geq \frac{\gamma}{\gamma+1}.$
    \item If $g_\infty  \in \big(\frac{1}{\gamma+1},\frac{\gamma}{\gamma+1}\big)$, then $\Phi = g_\infty$.
\end{enumerate}
\end{Theorem}

\thmref{Thm: global estimation} is established in \appref{Proof of Theorem global estimation}. It demonstrates that if the estimator \eqref{Eq: Estimator_general} converges to a value strictly between $(\gamma+1)^{-1}$ and $\gamma(\gamma+1)^{-1}$, then this value is actually equal to $\Phi$. In other words, our estimator is consistent provided that the population does not include large majorities. On the other hand, if the estimator converges to $(\gamma+1)^{-1}$ or $\gamma(\gamma+1)^{-1}$, then it cannot estimate the true value of $\Phi$. Instead, it gives a lower bound on the size of the majority. In this case, the inconsistency of the estimator \eqref{Eq: Estimator_general} is due to the presence of a large majority that forces the minority to lie with probability one in the limit. Indeed, $g_\infty = (\gamma+1)^{-1}$ (respectively $g_\infty = \gamma (\gamma+1)^{-1}$) implies that the majority of agents have inherent opinions equal to $0$ (respectively $1$), which corresponds to $(p_\infty^0,p_\infty^1) = (1,0)$ (respectively $(0,1)$). Hence, in the limit, the minority of agents with inherent opinion $1$ (respectively $0$) lie with probability one.  

In \lemref{lemma: individual convergence}, \eqref{Eq: partition} partitions the agents into two mutually exclusive groups. The first includes agents in $V^1$ whose average declared opinions over time  $\frac{1}{n} \sum_{k=0}^{n-1}\psi_{v,k}$ converge $a.s.$ to $p_{\infty}^1$, and the second contains agents in $V^0$ whose average declared opinions converge to $1-p_{\infty}^0$. (Recall that agents in $V^j$ have inherent opinion $j \in \{0,1\}$.) If the population does not include a large majority, then we can estimate $\Phi$ accurately. As a result, $p_\infty^0$ and $p_\infty^1$ can be computed explicitly. This means that the individual average declared opinions $\frac{1}{n} \sum_{k=0}^{n-1}\psi_{v,k}$ and the global average declared opinion $\hat \Psi_n$ can be used to estimate the individual inherent opinions $\{\phi_v : v\in V\}$. This is presented formally in the following corollary, which follows immediately from Theorem \ref{Thm: global estimation} and Lemma \ref{lemma: individual convergence}.

\begin{Corollary}[Individual Opinions II] \label{Cor: individual estimation}
Suppose that $\hat \Psi_\infty \in (0,1)$. Then, we have for every $v \in V$:
\begin{enumerate}
    \item If $\psi_{v,\infty} = \frac{\gamma(\gamma\Phi-(1-\Phi))}{\Phi(\gamma^2-1)}$, then $\phi_v = 1$,
    \item If $\psi_{v,\infty} = 1- \frac{\gamma(\gamma(1-\Phi)-\Phi)}{(1-\Phi)(\gamma^2-1)}$, then $\phi_v = 0$,
\end{enumerate}
where $\psi_{v,\infty}$ is given by \eqref{Eq: partition}.
\end{Corollary}

\section{Conclusion and Future Work} \label{sec:conclusion}

In this paper, we introduced a new interacting P\'{o}lya urn opinion dynamics model where a population of unreliable agents holds both inherent and declared opinions. The inherent opinions are static and unknown, while the possibly untrue declared opinions are dynamic and broadcast over a social network. In the special case of a complete graph, we propose an estimator to learn the agents' inherent opinions by observing their declared ones, and prove that our estimator is consistent provided that the population does not include large majorities. Furthermore, when a large majority exists, our estimator gives a lower bound on the size of the majority. It should be noted that our results only hold when agents have symmetric and homogeneous social influence  relationships that are known to the estimator, e.g., our estimator has knowledge of the honesty parameter $\gamma$, which is the same for all agents. Due to this, and the other limitations discussed in \secref{sec:model}, our results obviously do not capture the complexity of real-world opinion dynamics systems. However, this work initiates a rigorous mathematical study of the broader question of determining the true opinions of agents from their possibly fallacious declarations.

The aforementioned shortcomings of our model beget several useful directions of future research. Firstly, one could try to generalize our results to larger (non-binary) opinion  alphabets or to other graph topologies, e.g., expander graphs, random regular graphs, etc. Although extending our results beyond complete graphs to more realistic graph topologies is an important future direction, it will involve addressing a key technical difficulty. Indeed, much of the theory of stochastic approximations, e.g., the crucial results of \cite{renlund2011recursive}, has been developed for one-dimensional (1D) stochastic processes, and the martingale arguments used in such analysis are challenging to extend to multi-dimensional settings. Likewise, analyzing the stability of multi-dimensional ODEs (which might be engendered by multi-dimensional stochastic processes) is also more difficult than 1D ODEs. This effectively restricts us to using mainly 1D stochastic approximation arguments. In this work, focusing on complete graphs reduces our analysis of the opinion dynamics of all agents to studying the evolution of a single 1D process and its corresponding ODE. (Note that even this simplification relies on a nontrivial ``projection argument,'' where we define the 1D process $\tilde{X}_n$ in \appref{Proof of Theorem convergence}.) However, if one tries to generalize our results to other graphs, the agents are no longer symmetric, and it is not straightforward to transform the multi-dimensional opinion dynamics of all agents to a 1D stochastic process. This is the main obstacle that must be circumvented to successfully extend our results. Secondly, one could also try to develop an estimator for the honesty parameter $\gamma$, which is assumed to be known in our current model. Finally, since asymptotic estimation in the presence of large majorities appears to be impossible, one could explore learning from transient epochs of the stochastic process using concentration of measure ideas as $N \rightarrow \infty$.

\appendices

\section{Proof of \propref{Prop: Martingale Property}}
\label{Proof of Proposition Martingale Property}

\underline{Part 1:} To prove the first part, observe that we have 
\begin{align}
&\E\!\left[1-p_n \middle| \F_{n-1} \right]  = \E\!\left[\frac{\beta_n}{\alpha_n + \beta_n} \, \middle| \, \alpha_{n-1}, \beta_{n-1} \right] \nonumber \\
& = \E\!\left[\frac{\beta_{n-1} + \Psi_{n-1}}{\alpha_{n-1} + \beta_{n-1} + \gamma(1-\Psi_{n-1}) + \Psi_{n-1}} \, \middle| \, \alpha_{n-1}, \beta_{n-1} \right] \nonumber \\
& \leq \E\!\left[\frac{\beta_{n-1} + \Psi_{n-1}}{\alpha_{n-1} + \beta_{n-1} + 1} \, \middle| \, \alpha_{n-1}, \beta_{n-1} \right] \label{Eq: Inequality in Supermartingale} \\
& = \frac{\beta_{n-1} + \E\!\left[ \Psi_{n-1}  \middle| \alpha_{n-1}, \beta_{n-1} \right]}{\alpha_{n-1} + \beta_{n-1} + 1} \nonumber \\
& = \frac{\beta_{n-1} + 1 - p_{n-1}}{\alpha_{n-1} + \beta_{n-1} + 1} = \frac{(\alpha_{n-1}+\beta_{n-1})(1-p_n)+1-p_n}{\alpha_{n-1}+\beta_{n-1}+1} \nonumber \\
& =  1 - p_{n-1} \nonumber
\label{Eq: Super-Martingale Condition}
\end{align}
where the first equality follows from the Markovian nature of the dynamics \eqref{Eq: Dynamics of alpha}, \eqref{Eq: Dynamics of beta}, and \eqref{Eq: Modified Transition of Psi_t}, the second equality follows from \eqref{Eq: Dynamics of alpha} and \eqref{Eq: Dynamics of beta}, the third inequality holds because $(\gamma - 1)(1-\Psi_{n-1}) \geq 0$ since $\gamma > 1$, and the fifth equality holds because  $N \Psi_{n-1} \sim \bin(N,1-p_{n-1})$ given $\alpha_{n-1}$ and $\beta_{n-1}$. This establishes part 1. (Note that \eqref{Eq: Inequality in Supermartingale} is met with equality if $\gamma = 1$, which turns $\{p_n : n \in \Z_+\}$ into a martingale.)

\underline{Part 2:} The second part follows from the \emph{martingale convergence theorem} \cite[Chapter 4, Theorem 4.1]{Cinlar2011}.

\section{Proof of \thmref{Thm: Convergence to Bernoulli Variable}} 
\label{Proof of Theorem Convergence to Bernoulli Variable}

\underline{Part 1:} We will prove that $p^* = 1$ $a.s.$ using stochastic approximations in \thmref{Thm: Asymptotic Learning}, but for now, we present a more elementary proof of part 1 that will be useful in proving parts 2 and 3. First, by unwinding the recursions in \eqref{Eq: Dynamics of alpha} and \eqref{Eq: Dynamics of beta}, we obtain
\begin{align*}
\alpha_n & = \frac{1}{N} + \gamma n \big(1 - \hat{\Psi}_n\big) , \\
\beta_n & = \frac{1}{N} + n \hat{\Psi}_n \, ,
\end{align*}
for all $n \geq 1$, which implies that
$$ 1 - p_n = \frac{\frac{1}{N} + n \hat{\Psi}_n}{\frac{2}{N} + n \hat{\Psi}_n  + \gamma n \big(1 - \hat{\Psi}_n\big)} = \frac{\frac{1}{N n} + \hat{\Psi}_n}{\frac{2}{Nn} + \gamma - (\gamma - 1) \hat{\Psi}_n} . $$
Then, letting $n \rightarrow \infty$, we get using part 2 of \propref{Prop: Martingale Property} that
$$ 1 - p^* = \lim_{n \rightarrow \infty}{\frac{\hat{\Psi}_n}{\gamma - (\gamma - 1) \hat{\Psi}_n}} \enspace a.s. $$
which, after applying the function $g:[0,1] \rightarrow [0,1]$, $g(x) = \gamma x/(1 + (\gamma-1)x)$, yields
\begin{equation}
\label{Eq: Convergence of Mean of Psi}
\begin{split}
\lim_{n \rightarrow \infty}{\hat{\Psi}_n} &= g\!\left(\lim_{n \rightarrow \infty}{\frac{\hat{\Psi}_n}{\gamma - (\gamma - 1) \hat{\Psi}_n}}\right)\\
&= g(1 - p^*) =  \frac{\gamma (1 - p^*)}{1 + (\gamma-1)(1 - p^*)} \enspace a.s.
\end{split}
\end{equation}
where we use the continuity of $g$ (along with the Mann-Wald continuous mapping theorem). Hence, using Lebesgue's \emph{dominated convergence theorem} \cite[Chapter 1, Theorem 4.16]{Cinlar2011}, we have
\begin{equation}
\label{Eq: Mean Form 1}
\lim_{n \rightarrow \infty}{\E\!\left[\hat{\Psi}_n\right]} = \E\!\left[g(1 - p^*)\right] .
\end{equation}
On the other hand, since $N \Psi_n \sim \bin(N,1-p_n)$ given $p_n$ for all $n \in \Z_+$, we have
$$ \lim_{n \rightarrow \infty}{\E\!\left[\Psi_n\right]} = 1-\lim_{n \rightarrow \infty}{\E\!\left[p_n\right]} = 1 - \E\!\left[p^*\right] , $$
where we utilize \eqref{Eq: Expected Value Bounds} (which also follows from \propref{Prop: Martingale Property}). Then, using \eqref{Eq: Estimator Definition} and Ces\`{a}ro mean convergence \cite[Chapter 3, Exercise 14(a)]{Rudin1976}, we obtain
\begin{equation}
\label{Eq: Mean Form 2}
\lim_{n \rightarrow \infty}{\E\!\left[\hat{\Psi}_n\right]} = \lim_{n\rightarrow \infty}{\frac{1}{n}\sum_{k = 0}^{n-1}{\E\!\left[\Psi_k\right]}} = 1 - \E\!\left[p^*\right] .
\end{equation}
Combining \eqref{Eq: Mean Form 1} and \eqref{Eq: Mean Form 2} produces the key equation:
\begin{equation}
\label{Eq: Key Equation}
\E\!\left[1 - p^*\right] = \E\!\left[g(1-p^*)\right] .
\end{equation}

Now suppose for the sake of contradiction that $\P(p^* \in (0,1)) > 0$. To prove part 1, notice that $g$ is a strictly increasing and strictly concave function on $[0,1]$ (since $\gamma > 1$); indeed, $g^{\prime}(x) = \gamma/(1 + (\gamma-1)x)^2 > 0$ and $g^{\prime \prime}(x) = -2\gamma(\gamma-1)/(1 + (\gamma - 1)x)^3 < 0$ for all $x \in [0,1]$. Since $g(0) = 0$ and $g(1) = 1$, we must have $g(x) > x$ for all $x \in (0,1)$. Let us rewrite \eqref{Eq: Key Equation} as follows:
\begin{align*}
&\underbrace{\P\!\left(p^* = 0\right) + \E\!\left[1 - p^* \, \middle| \, p^* \in (0,1)\right] \P(p^* \in (0,1))}_{= \, \E\!\left[1 - p^*\right]}\\
&= \underbrace{\P\!\left(p^* = 0\right) + \E\!\left[g(1-p^*) \, \middle| \, p^* \in (0,1)\right] \P(p^* \in (0,1))}_{= \, \E\!\left[g(1 - p^*)\right]} 
\end{align*}
which implies that 
$$ \E\!\left[1 - p^* \, \middle| \, p^* \in (0,1)\right] = \E\!\left[g(1-p^*) \, \middle| \, p^* \in (0,1)\right] $$
where the conditional expectations are well-defined because $\P(p^* \in (0,1)) > 0$ (by assumption). Since $g(x) > x$ for all $x \in (0,1)$ and $\P(1-p^* \in (0,1)) > 0$, we must also have (cf. \cite[Chapter 1, Exercises 4.25 and 4.32]{Cinlar2011})
$$ \E\!\left[1 - p^* \, \middle| \, p^* \in (0,1)\right] < \E\!\left[g(1-p^*) \, \middle| \, p^* \in (0,1)\right] , $$
which is a contradiction. Thus, we have established that $\P(p^* \in (0,1)) = 0$, or equivalently, $\P(p^* \in \{0,1\}) = 1$. This proves part 1.

\underline{Part 2:} To prove part 2, recall from \eqref{Eq: Convergence of Mean of Psi} that $\lim_{n \rightarrow \infty}{\hat{\Psi}_n} = g(1 - p^*)$ $a.s.$ Since $1-p^* \in \{0,1\}$ $a.s.$ due to part 1, and $g(0) = 0$ and $g(1) = 1$, we have $\lim_{n \rightarrow \infty}{\hat{\Psi}_n} = 1 - p^*$ $a.s.$ This establishes part 2.

\underline{Part 3:} Finally, to prove part 3, observe that for all $k \in \{0,\dots,N\}$,
\begin{align*}
\lim_{n \rightarrow \infty}{\P\!\left(N \Psi_n = k \, \middle| \, p_n\right)} &= \lim_{n \rightarrow \infty}{\binom{N}{k} (1 - p_n)^{k} p_n^{N-k}} \\
&= \binom{N}{k} (1 - p^*)^{k} (p^*)^{N-k}, \end{align*}
where we use the fact that $N \Psi_n \sim \bin(N,1-p_n)$ given $p_n$ for all $n \in \Z_+$. Using the dominated convergence theorem, this produces
\begin{align*}
\lim_{n \rightarrow \infty}{\P\!\left(\Psi_n = \frac{k}{N}\right)} & = \binom{N}{k} \E\!\left[(1 - p^*)^{k} (p^*)^{N-k}\right] \\
& = \begin{cases}
\P\!\left(p^* = 1\right) , & k = 0 \\
\P\!\left(p^* = 0\right) , & k = N \\
0 \, , & k \in \{1,\dots,N-1\}
\end{cases}
\end{align*}
where the second equality follows from part 1. Therefore, $\Psi_n$ converges weakly (or in distribution) to $1-p^*$ as $n \rightarrow \infty$. This completes the proof.

\section{Proof of \propref{Prop: Non-existence of Almost Sure Limit}}
\label{Proof of Proposition Non-existence of Almost Sure Limit}

It suffices to prove that $\P(\lim_{n \rightarrow \infty}{\Psi_n} \in \{0,1\}) = 0$, since using part 1 of \thmref{Thm: Convergence to Bernoulli Variable}, this implies that $\{\Psi_n : n \in \Z_+\}$ does not converge to $1-p^*$ $a.s.$ as $n \rightarrow \infty$. To this end, for any $n_0 \in \Z_+$, we first define the quantities:
$$ \forall n \geq n_0, \enspace \Lambda_{n_0:n} \triangleq \P\!\left(\forall n_0 \leq  s \leq n, \, \Psi_s = 1\right) . $$
Observe that for every $n \geq n_0$, this quantity satisfies the upper bound:
\begin{align*}
&\Lambda_{n_0:n}  = \P\!\left(\Psi_n = 1 \, \middle| \, \forall n_0 \leq s \leq n-1, \, \Psi_s = 1\right)
\Lambda_{n_0:n-1} \\
& = \E\!\left[\P\!\left(\Psi_n = 1 \, \middle| \, \alpha_n, \beta_n\right) \, \middle| \, \forall n_0 \leq s \leq n-1, \, \Psi_s = 1\right]
\Lambda_{n_0:n-1} \\
& = \E\!\left[\left(\frac{\beta_n}{\alpha_n + \beta_n}\right)^{\! N} \, \middle| \, \forall n_0 \leq s \leq n-1, \, \Psi_s = 1\right] \Lambda_{n_0:n-1} \\
& \leq \left(\frac{\frac{1}{N} + n}{\frac{2}{N} + n}\right)^{\! N} \Lambda_{n_0:n-1} \\
& \leq \left(\prod_{k = n_0}^{n}{\frac{\frac{1}{N} + k}{\frac{2}{N} + k}}\right)^{\! N}
\end{align*}
where the second and third equalities hold because $N \Psi_n \sim \bin(N,1-p_n)$ is conditionally independent of $\{\Psi_s : n_0 \leq s \leq n-1\}$ given $\alpha_n$ and $\beta_n$, the fourth inequality holds because $\beta_n \leq \beta_0 + n$ (using \eqref{Eq: Dynamics of beta}) and $\alpha_n + \beta_n \geq \alpha_0 + \beta_0 + n$ (using \eqref{Eq: Dynamics of alpha}, \eqref{Eq: Dynamics of beta}, and the fact that $\gamma > 1$), and the fifth inequality follows from unwinding the recursion. Letting $n \rightarrow \infty$, we obtain that
\begin{align}
\lim_{n \rightarrow \infty}{\Lambda_{n_0:n}} & \leq \left(\lim_{n \rightarrow \infty}{\prod_{k = n_0}^{n}{\frac{\frac{1}{N} + k}{\frac{2}{N} + k}}}\right)^{\! N} \nonumber \\
& = \left(\exp\!\left(\sum_{k = n_0}^{\infty}{\log\!\left(\frac{\frac{1}{N} + k}{\frac{2}{N} + k}\right)}\right)\right)^{\! N} \nonumber \\
& = \left(\exp\!\left(-\sum_{k = n_0}^{\infty}{\log\!\left(1 + \frac{1}{1 + N k}\right)}\right)\right)^{\! N} ,
\label{Eq: Intermediate Bound}
\end{align}
where the first inequality and the second equality use the continuity of the monomial function $x \mapsto x^n$ and the natural logarithm function $x \mapsto \log(x)$, respectively. Now notice that since $\lim_{k \rightarrow \infty}{\log(1 + (1+Nk)^{-1})/(1+Nk)^{-1}} = \lim_{x \rightarrow 0}{\log(1 + x)/x} = 1$ (by L'H\^{o}pital's rule \cite[Theorem 5.13]{Rudin1976}) and $\sum_{k = n_0}^{\infty}{(1 + N k)^{-1}} = +\infty$ (since the harmonic series diverges \cite[Theorem 3.28]{Rudin1976}), we have
$$ \sum_{k = n_0}^{\infty}{\log\!\left(1 + \frac{1}{1 + N k}\right)} = +\infty $$
using the limit comparison test \cite[Theorem 8.21]{Apostol1974}. Using \eqref{Eq: Intermediate Bound}, this yields
$$ \lim_{n \rightarrow \infty}{\Lambda_{n_0:n}} = \lim_{n \rightarrow \infty}{\P\!\left(\forall n_0 \leq  s \leq n, \, \Psi_s = 1\right)} = 0 $$
which, via the continuity of $\P$, implies that
$$ \P\!\left(\forall s \geq n_0, \, \Psi_s = 1\right) = 0 $$
because $\{\{\forall n_0 \leq  s \leq n, \, \Psi_s = 1\} : n \geq n_0\}$ is a non-increasing sequence of events with limit $$\bigcap_{n \geq n_0}{\{\forall n_0 \leq  s \leq n, \, \Psi_s = 1\}} = \{\forall s \geq n_0, \, \Psi_s = 1\}.$$ Then, since the above equality holds for all $n_0 \in \Z_+$, applying the union bound, we get
\begin{equation}
\label{Eq: Limit is not 1}
\P\!\left(\lim_{s \rightarrow \infty}{\Psi_s} = 1\right) = \P\!\left(\exists n_0 \in \Z_+, \, \forall s \geq n_0, \, \Psi_s = 1\right) = 0 \, , 
\end{equation}
where the first equality holds because $N \Psi_s \in \{0,\dots,N\}$ for all $s \in \Z_+$, which means that $\lim_{s \rightarrow \infty}{\Psi_s} = 1$ if and only if there exists $n_0 \in \Z_+$ such that for all $s \geq n_0$, $\Psi_s = 1$. Similarly, since $p_n \leq (\alpha_0 + \gamma n)/(\alpha_0 + \beta_0 + \gamma n)$ (using \eqref{Eq: Dynamics of alpha} and \eqref{Eq: Dynamics of beta}), we can show that
\begin{equation}
\label{Eq: Limit is not 0}
\P\!\left(\lim_{s \rightarrow \infty}{\Psi_s} = 0\right) = 0 \, . 
\end{equation}
By the union bound, \eqref{Eq: Limit is not 1} and \eqref{Eq: Limit is not 0} yield $\P(\lim_{n \rightarrow \infty}{\Psi_n} \in \{0,1\}) = 0$ as desired. This completes the proof.

\section{Proof of \thmref{Thm: Asymptotic Learning}}
\label{Proof of Theorem Asymptotic Learning}

\underline{Part 1:}  To prove this, we employ the powerful technique of stochastic approximation. Observe that for any $n \in \Z_+$,
\begin{align*}
    & p_{n+1}-p_n  = \frac{\alpha_{n}+\gamma (1-\Psi_n)}{\alpha_n + \beta_n + \gamma (1-\Psi_n) + \Psi_n} - \frac{\alpha_n}{\alpha_n + \beta_n} \\
    & = \frac{\alpha_{n}+\gamma (1-\Psi_n) - \frac{\alpha_n}{\alpha_n + \beta_n} (\alpha_n + \beta_n + \gamma (1-\Psi_n) + \Psi_n)}{\alpha_n + \beta_n + \gamma (1-\Psi_n) + \Psi_n}  \\
    & = \frac{\gamma (1-\Psi_n) - p_n ( \gamma (1-\Psi_n) + \Psi_n) }{ \alpha_n + \beta_n + \gamma (1-\Psi_n) + \Psi_n} \\
    &= \gamma_{n+1}  \left( F(p_n) + \xi_{n+1} \right),
\end{align*}
where we let
\begin{align*}
F(p) &= -(\gamma -1) p^2 + (\gamma   -1 ) p, \enspace \text{for } p \in [0,1] , \\
\xi_{n+1} &= \gamma (1-\Psi_n) - p_n ( \gamma (1-\Psi_n) + \Psi_n) - F(p_n) \\
&=(\gamma(1-p_n)+p_n)(1-\Psi_n-p_n) , \\
\gamma_{n+1}&=\frac{1}{ \alpha_n + \beta_n + \gamma (1-\Psi_n) + \Psi_n} ,
\end{align*}
and we have $\frac{1}{(2/N)+ \gamma (n+1)}\leq \gamma_{n+1} \leq  \frac{1}{(2/N) + 1 + n}$. Note that we decompose $p_{n+1} - p_n$ in this way in order to employ \cite[Theorems 1 and 3]{renlund2011recursive} in the sequel.

Next, we show that $\E[\gamma_{n+1}\xi_{n+1}|\F_{n}] \leq C/n^2$ for some constant $C>0$. For all sufficiently large $n \in \Z_+$, 
\begin{align*}
&\E\!\left[\gamma_{n+1} \xi_{n+1} \middle| \F_{n}\right]  \\
&= \E\!\left[\frac{\gamma(1 - p_n) - F(p_n) + (p_n(\gamma - 1) - \gamma) \Psi_n}{\alpha_n + \beta_n + \gamma - (\gamma - 1)\Psi_n} \middle| \F_{n}\right] \\
& = \left(\frac{\gamma}{\gamma - 1} - p_n\right) + \left(\frac{\gamma(1-p_n) - F(p_n)}{\alpha_n + \beta_n + \gamma} - \frac{\gamma}{\gamma - 1} + p_n\right)  \\
&\times \E\!\left[\frac{1}{1 - \left(\frac{\gamma - 1}{\alpha_n + \beta_n + \gamma}\right)\Psi_n} \middle| \F_{n}\right] \\
& = \left(\frac{\gamma}{\gamma - 1} - p_n\right) + \left(\frac{\gamma(1-p_n) - F(p_n)}{\alpha_n + \beta_n + \gamma} - \frac{\gamma}{\gamma - 1} + p_n\right) \\ & \times \sum_{k = 0}^{\infty}{\E\!\left[\left(\frac{\gamma - 1}{\alpha_n + \beta_n + \gamma}\right)^{\! k} \Psi_n^k \middle| \F_{n}\right]} \\
& = \left(\frac{\gamma}{\gamma - 1} - p_n\right) \!+\! \left(\frac{\gamma(1-p_n) - F(p_n)}{\alpha_n + \beta_n + \gamma} - \frac{\gamma}{\gamma - 1} + p_n\right) \! \times\\ &  \left(\!1 + \frac{(\gamma - 1)(1 - p_n)}{\alpha_n + \beta_n + \gamma} + \sum_{k = 2}^{\infty}{\E\!\left[\left(\frac{\gamma - 1}{\alpha_n + \beta_n + \gamma}\right)^{\! k} \Psi_n^k \middle| \F_{n}\right]}\right) \\
& = \left(\frac{\gamma(1-p_n) - F(p_n)}{\alpha_n + \beta_n + \gamma} \right) \! \left(\frac{(\gamma - 1)(1 - p_n)}{\alpha_n + \beta_n + \gamma} \right) \\ & + \left(\frac{\gamma(1-p_n) - F(p_n)}{\alpha_n + \beta_n + \gamma} - \frac{\gamma}{\gamma - 1} + p_n\right) \\
& \times \sum_{k = 2}^{\infty}{\E\!\left[\left(\frac{\gamma - 1}{\alpha_n + \beta_n + \gamma}\right)^{\! k} \Psi_n^k \middle| \F_{n}\right]}\leq  \frac{C}{n^2},
\end{align*}
where the second equality follows from a partial fraction decomposition, the third equality follows from the geometric series formula and Tonelli's theorem, the fourth equality holds because $N \Psi_n \sim \bin(N,1-p_n)$ given $\F_n$, the fifth equality follows from substituting the expression for $F$, and the final inequality holds because $\alpha_n + \beta_n \geq n$, where $C>0$ depends only on $\gamma$. By \cite[Theorem 1]{renlund2011recursive}, we can deduce that $p_n$ converges $a.s.$ to $0$ or $1$ (cf. part 1 of \thmref{Thm: Convergence to Bernoulli Variable}). 

In the following, we show  using \cite[Theorem 3]{renlund2011recursive} that $p_n$ does not converge $a.s.$ to $0$. Observe that $F(x)x>0$ for all $x\in (0,1)$, i.e., $0$ is an unstable point of the ODE defined by the vector field $F$. We have
$F(p)^2=(\gamma-1)^2p^2(1-p)^2 \leq (\gamma-1)^2 p$.
Moreover, using \eqref{Eq: Modified Transition of Psi_t}, we get 
$$\E\!\left[\xi_{n+1}^2 \middle| \F_n\right]=\  \frac{(\gamma(1-p_n)+p_n )^2}{N}p_n(1-p_n)\leq \frac{\gamma^2}{N}p_n.$$ 
By Proposition \ref{Prop: Non-existence of Almost Sure Limit}, we obtain that $\alpha_n\to \infty$ $a.s.$ as $n\to \infty$. Furthermore, $$np_n=\frac{n}{\alpha_n+\beta_n}\alpha_n \geq \frac{n}{(2/N) +\gamma n}\alpha_n.$$
Hence, $\lim_{n \rightarrow \infty}{np_n} = \infty$ $a.s.$ Applying 
\cite[Theorem 3]{renlund2011recursive}, we obtain that $\P(\lim_{n \rightarrow \infty}{p_n} = 0)=0$. This proves the first part.

\underline{Part 2:} The second part follows from part 1 and part 2 of \thmref{Thm: Convergence to Bernoulli Variable}, which shows that $\hat{\Psi}_n \rightarrow 0$ $a.s.$ as $n \rightarrow \infty$.

\section{Proof of \lemref{lem: domains}}
\label{Proof of Lemma domains}

\underline{Part 1:}
The first part follows from the definitions of the processes $X_n$, $Y_n$, and $Z_n$ given by \eqref{Eq: definition X}-\eqref{Eq: definition Z}. 

\underline{Part 2:} It is sufficient to show that the vector field $G$ is pointing to the interior of the manifold $M$ at the boundary $\partial M$ of $M$, i.e. $\langle G(x,y,z),n(x,y,z) \rangle \geq 0$ for all $(x,y,z) \in \partial M$, where  $n(x,y,z)$ is a normal vector to $\partial M$ at $(x,y,z)$ pointing to the interior of $M$, and $\langle.,.\rangle$ is the standard inner product on $\R^3$. At the boundary $x+y=\frac{1}{\gamma+1}$, we have
\begin{align*}
    &\langle G(x,y,z),n(x,y,z)\rangle=
\langle G(x,y,z),(1,1,0)\rangle \\
&=-(\gamma+1)(x+y)\\
&+\gamma\left( (1-\Phi) \frac{x}{x+y} + \Phi \left( 1- \frac{z}{1-x-y} \right) \right)  \\
&+  (1-\Phi) \left(1-\frac{x}{x+y} \right) + \Phi \frac{z}{1-x-y} \\
& = -1  +\gamma\left( (1-\Phi) \frac{x}{x+y} + \Phi \left( 1- \frac{z}{1-x-y} \right) \right)  \\
&+  (1-\Phi) \left(1-\frac{x}{x+y} \right) + \Phi \frac{z}{1-x-y} \\
& \geq  -1  +\left( (1-\Phi) \frac{x}{x+y} + \Phi \left( 1- \frac{z}{1-x-y} \right) \right)  \\
&+  (1-\Phi) \left(1-\frac{x}{x+y} \right) + \Phi \frac{z}{1-x-y} =0. 
\end{align*}
At the boundary $x+y=\frac{\gamma}{\gamma+1}$, we have
\begin{align*}
    &\langle G(x,y,z),n(x,y,z)\rangle=
\langle G(x,y,z),(-1,-1,0)\rangle \\
&=(\gamma+1)(x+y) \\
&-\gamma\left( (1-\Phi) \frac{x}{x+y} + \Phi \left( 1- \frac{z}{1-x-y} \right) \right) \\
&-  (1-\Phi) \left(1-\frac{x}{x+y} \right) + \Phi \frac{z}{1-x-y} \\
& =\gamma   -\gamma\left( (1-\Phi) \frac{x}{x+y} + \Phi \left( 1- \frac{z}{1-x-y} \right) \right) \\
& -  (1-\Phi) \left(1-\frac{x}{x+y} \right) + \Phi \frac{z}{1-x-y} \\
& \geq  \gamma -  \gamma \left( (1-\Phi) \frac{x}{x+y} + \Phi \left( 1- \frac{z}{1-x-y} \right) \right)  \\
&-\gamma   (1-\Phi) \left(1-\frac{x}{x+y} \right) + \Phi \frac{z}{1-x-y} =0. 
\end{align*}
At the boundary $x=0$, we have
\begin{align*}
    \langle G(x,y,z),n(x,y,z)\rangle&=
\langle G(x,y,z),(1,0,0)\rangle \\
&= \gamma \Phi \left( 1- \frac{z}{1-y} \right) \geq 0. 
\end{align*}
At the boundary $x=\frac{\gamma}{\gamma+1}$, we have
\begin{align*}
    &\langle G(x,y,z),n(x,y,z)\rangle=
\langle G(x,y,z),(-1,0,0)\rangle \\
&= 
\gamma - \gamma \left( (1-\Phi) \frac{x}{x+y} + \Phi \left( 1- \frac{z}{1-x-y} \right) \right) \geq 0.
\end{align*}
The proof for the rest of the boundaries is analogous.

\underline{Part 3:} The third part follows from the fact that the vector field $G$ is smooth on the compact invariant set $M$. 

\section{Proof of \lemref{lem: equilrium points}}
\label{Proof of Lemma equilrium points}

\subsection{Preliminary Definitions}

The following definitions will be used in the proof.

\begin{Definition}[Semiflow \cite{benaim1999dynamics}] \label{def:demiflow}
A semiflow $\Phi$ on a smooth manifold $M$ is a continuous map 
\begin{align*}
    \Phi  : \R_+\times M &\to M \\
    (t,x) & \mapsto \Phi(t,x) = \Phi_t(x),
\end{align*}
such that $\Phi_0(x)=x$ and $\Phi_{t+s}=\Phi_t \circ \Phi_s$. 
\end{Definition}

\begin{Definition}[Pseudotrajectory \cite{benaim1999dynamics}] \label{def:pseudotrajectory}
A continuous function $X: \R_+ \to M$ is an asymptotic pseudotrajectory for a semiflow $\Phi$ if 
\[\forall T>0, \enspace \lim_{t \to \infty} \sup\limits_{0 \leq h \leq T} \|X(t+h)-\Phi_h(X(t))\|=0 \, .
\]
\end{Definition}

\begin{Definition}[Transitivity \cite{benaim1999dynamics}] \label{def:transitive}
A compact invariant set $L$ of a semiflow $\Phi$ on $M$ is internally chain transitive if for all $\delta>0$ and $T>0$, for all
$a,b\in L$, there exists $y_0,\dots,y_k \in L$ such that $\|a-y_0\|\leq \delta$, $\|\Phi_{t_j}(y_j)-y_{j+1}\|\leq \delta$, for $j=1,\dots,k-1$, and $y_k=b$. (Here, invariant means that $\Phi_t(L)=L$ for all $t \geq 0$.)
\end{Definition}

\subsection{Proof of Lemma \ref{lem: equilrium points}}

\underline{Part 1:} Under the change of variables $d=x+\gamma y$ and $e=x+z$, the ODE \eqref{Eq: ODE} is equivalent to the following ODE:
\begin{align} \label{Eq: ODE1}
 \frac{d}{dt}\begin{pmatrix}
   x \\ d \\e
   \end{pmatrix} = \tilde G(x,d,e) \triangleq  
   \begin{pmatrix}
    g_1(x) + g_2(x,d,e)\\ -(\gamma+1)d+\gamma \\  -(\gamma+1)e+\gamma 
   \end{pmatrix},
\end{align}
where 
\begin{align}
&g_1(x)  = \frac{(\gamma+1)(\gamma-1)^2}{\left((\gamma-1)x+\frac{\gamma}{\gamma+1}\right)\left(\frac{\gamma^2}{\gamma+1}-(\gamma-1)x\right)} \times  \label{Eq: g_1}\\
&x\left(x-\frac{\gamma}{\gamma+1}\right)\left(x-\frac{\gamma}{\gamma^2-1}(\gamma(1-\Phi)-\Phi)\right)\nonumber \\
&g_2(x,d,e)  = G^1\left (x, \frac{1}{\gamma}(d-x),e-x\right ) \label{Eq: g_2} \\
&-G^1\left (x, \frac{1}{\gamma}\left(\frac{\gamma}{\gamma+1}-x\right),\frac{\gamma}{\gamma+1}-x\right )\nonumber \\   &=\frac{(1-\Phi)\gamma^2 x\left (\frac{\gamma}{\gamma+1}-d\right)}{\left((\gamma-1)x+\frac{\gamma}{\gamma+1}\right)\left((\gamma-1)x+d\right)} \nonumber \\
&+   \frac{\Phi \gamma x\left (d-\frac{\gamma}{\gamma+1}\right)}{\left (\frac{\gamma^2}{\gamma+1}-(\gamma-1)x\right)(\gamma-d-(\gamma-1)x)} \nonumber \\
&  -\frac{\left (\frac{\gamma^2}{\gamma+1}-(\gamma-1)x\right )\left (\gamma \left (e-\frac{\gamma}{\gamma+1}\right)-\left(d-\frac{\gamma}{\gamma+1}\right )\right )}{\left (\frac{\gamma^2}{\gamma+1}-(\gamma-1)x\right)(\gamma-d-(\gamma-1)x)}. \nonumber 
\end{align}
$(x_0,y_0,z_0)$ is an equilibrium point of \eqref{Eq: ODE} if and only if the corresponding $(x_0,d_0,e_0)$ is an equilibrium point of \eqref{Eq: ODE1}. Solving for $\tilde G(x_0,d_0,e_0)=0$ gives
$d_0=e_0=\gamma/(\gamma+1)$ and $g_1(x_0)=0$.
Noting that   $\frac{\gamma}{\gamma^2-1}(\gamma(1-\Phi)-\Phi)\in[0,\frac{\gamma}{\gamma+1}]$ if and only if $\gamma \geq \max \left ( \frac{\Phi}{1-\Phi},\frac{1-\Phi}{\Phi} \right)$, we deduce that $g_1(x_0)=0$ if and only if $x_0=0$ or $\gamma/(\gamma+1)$ in case  
$\gamma < \max \left ( \frac{\Phi}{1-\Phi},\frac{1-\Phi}{\Phi} \right)$, and $x_0=0$, $\gamma/(\gamma+1)$, or $\frac{\gamma}{\gamma^2-1}(\gamma(1-\Phi)-\Phi)$, otherwise. The first part follows by computing the corresponding $(x_0,y_0,z_0)$.

\underline{Part 2:} Let $\Phi^g$ be the semiflow (see Definition \ref{def:demiflow}) generated by the vector field $(g_1(x),-(\gamma+1)d+\gamma,-(\gamma+1)e+\gamma)$. Let $(X(t),D(t),E(t))$ be a solution of \eqref{Eq: ODE1}. In the following, we show that $(X(t),D(t),E(t))$ is an asymptotic pseudotrajectory (see Definition \ref{def:pseudotrajectory}) for $\Phi^g$. Fix $T>0$ and let $h\in[0,T]$. Then, 
\begin{align*}
    &\left\|\left(X(t+h),D(t+h),E(t+h)\right)-\Phi^g_h\left(X(t),D(t),E(t)\right)\right\|  \\
    &=\Big\| \int_0^{h} \Big(g_1(X(t+\tau)) - g_1(\Tilde{X}(\tau))\\
    &+g_2\left(X(t+\tau),D(t+\tau),E(t+\tau)\right)\Big )d\tau  \Big\|\\
    & \leq k_1\int_0^h \Big\|\left(X(t+\tau),D(t+\tau),E(t+\tau)\right)\\
    &-\Phi^g_\tau\left(X(t),D(t),E(t)\right)\Big\|d\tau +\\
    & k_2\int_0^h \left( \frac{1}{\gamma}\left\|D(t+\tau)-\frac{\gamma}{\gamma+1}\right\|+\left\|E(t+\tau)-\frac{\gamma}{\gamma+1}\right\| \right) d\tau\\
    & \leq k_1\int_0^h \Big\|\left(X(t+\tau),D(t+\tau),E(t+\tau)\right)\\
    & -\Phi^g_\tau\left(X(t),D(t),E(t)\right)\Big\|d\tau \\
    &+ k_2\left( \frac{1}{\gamma}D(0)+E(0) \right)\int_0^h e^{-(\gamma+1)(t+\tau)} d\tau ,
\end{align*}
where the first equality follows from the fact that $\Phi_\tau^g(X(t),D(t),E(t))=(\tilde X(\tau),D(\tau+t),E(\tau+t))$ where $\tilde X(\tau)$ is the unique solution of $dx/d\tau=g_1(x)$ with $x(t)=X(t)$, the second inequality follows from the Lipschitz continuity of $g_1$ and $G_1$ (where $k_1>0$ and $k_2>$ are the corresponding Lipschitz constants, respectively), and the final inequality follows from the fact that $D(\tau)$ and $E(\tau)$ are the unique solutions of $dx/d\tau=-(\gamma+1)x$ with initial conditions $D(0)$ and $E(0)$, respectively. By \emph{Gr{\"o}nwall's inequality} \cite{bellman1943stability}, we obtain
\begin{align*}
    &\left\|\left(X(t+h),D(t+h),E(t+h)\right)-\Phi^g_h\left(X(t),D(t),E(t)\right)\right\| \\
    &\leq  k_2\left( \frac{1}{\gamma}D(0)+E(0) \right)\int_0^h e^{-(\gamma+1)(t+\tau)} d\tau e^{k_1h} \\
    & \leq k_2\left( \frac{1}{\gamma}D(0)+E(0) \right) T e^{k_1T} e^{-(\gamma+1)t}.
\end{align*}
Hence, 
\begin{align}
    &\max\limits_{h\in[0,T]}\Big\|\left(X(t+h),D(t+h),E(t+h)\right)\nonumber \\
    &-\Phi^g_h\left(X(t),D(t),E(t)\right)\Big\| \nonumber \\
    &\leq k_2\left( \frac{1}{\gamma}D(0)+E(0) \right) T e^{k_1T} e^{-(\gamma+1)t}. \label{eq:pseudo}
\end{align}
This implies that the left hand side of \eqref{eq:pseudo} converges to $0$ as $t\to \infty$ and proves that $(X(t),D(t),E(t))$ is an asymptotic pseudotrajectory for $\Phi^g$. By \cite[Theorem 5.7]{benaim1999dynamics}, the limit points of $(X(t),D(t),E(t))$ is internally chain transitive of $\Phi^g$. But, the only internally chain transitive sets of $\Phi^g$ are 
its equilibrium points $\left(0,\frac{\gamma}{\gamma+1},\frac{\gamma}{\gamma+1}\right)$, $\left(\frac{\gamma}{\gamma+1},\frac{\gamma}{\gamma+1},\frac{\gamma}{\gamma+1}\right)$, and 
$\left(\frac{\gamma(\gamma(1-\Phi)-\Phi)}{\gamma^2-1},\frac{\gamma}{\gamma+1},\frac{\gamma}{\gamma+1}\right)$ if $\gamma \geq  \max\left( \frac{\Phi}{1-\Phi},\frac{1-\Phi}{\Phi}\right)$, or $\left(0,\frac{\gamma}{\gamma+1},\frac{\gamma}{\gamma+1}\right)$ and $\left(\frac{\gamma}{\gamma+1},\frac{\gamma}{\gamma+1},\frac{\gamma}{\gamma+1}\right)$, if $\gamma <  \max\left( \frac{\Phi}{1-\Phi},\frac{1-\Phi}{\Phi}\right).$

This was proved for any solution of \eqref{Eq: ODE1}. This implies that all the solutions of \eqref{Eq: ODE} converge to the equilibria of \eqref{Eq: ODE}.  Thus, the only internally chain sets of the flow generated by \eqref{Eq: ODE} are the equilibrium points of \eqref{Eq: ODE}. By  \cite[Proposition 4.1]{benaim1999dynamics} and \cite[Proposition 5.7]{benaim1999dynamics}, $(X_n,Y_n,Z_n)$ converges to an equilibrium point $(X_\infty,Y_\infty,Z_\infty)\in L$. Finally, part 2 follows from $p_n^0=X_n/(X_n+Y_n)$ and $p_n^1=Z_n/(1-X_n-Y_n)$. 

\section{Proof of Theorem \ref{Thm: convergence}}
\label{Proof of Theorem convergence}

We only prove the case $\gamma \geq  \max\left( \frac{\Phi}{1-\Phi},\frac{1-\Phi}{\Phi}\right)$. The proofs of the other cases are analogous. By \lemref{lem: equilrium points}, it is sufficient to show that $(p_n^0,p_n^1)$ does not converge to $(0,1)$ or $(1,0)$. Equivalently, it is sufficient to show that $X_n$ does not converge to $0$ or $\frac{\gamma}{\gamma+1}$. By defining the processes $D_n=X_n+\gamma Y_n$ and $E_n=X_n+Z_n$, the process $\tilde X_n=X_n+\frac{\gamma}{\gamma+1}-D_n+\frac{\gamma}{\gamma+1}-E_n$ satisfies the following equation: 
$$\tilde X_{n+1}-\tilde X_n=\gamma_{n+1} \left( g_1(\tilde X_n)+\tilde U^1_{n+1}\right),$$
where 
\begin{align*}
    \tilde U^1_{n+1}&=g_1(X_n)-g_1(\tilde X_n)+g_2(X_n,D_n,E_n)\\
    &-(\gamma+1)\left ( \frac{\gamma}{\gamma+1} -D_n +  \frac{\gamma}{\gamma+1} -E_n  \right )+U_{n+1}^1.
\end{align*}
As in the proof of \thmref{Thm: Asymptotic Learning}, note that we decompose $\tilde{X}_{n+1} - \tilde{X}_n$ in this way in order to apply \cite[Theorem 3]{renlund2011recursive} in the sequel.

The processes $D_n$ and $E_n$ can be computed explicitly as follows:
\begin{align}
   D_n-\frac{\gamma}{\gamma+1}&=\frac{\alpha_n^0+\gamma \beta_n^0}{\alpha_n^0+\beta_n^0+\alpha_n^1+\beta_n^1}-\frac{\gamma}{\gamma+1}\nonumber \\
   &=\frac{2/N+n\gamma}{4/N+n(\gamma+1)}-\frac{\gamma}{\gamma+1}\nonumber \\
   &=\frac{2/N(1-\gamma)}{4/N+n(\gamma+1)}, \label{Eq: D_n}\\
   E_n-\frac{\gamma}{\gamma+1}&=\frac{2/N+n\gamma}{4/N+n(\gamma+1)}-\frac{\gamma}{\gamma+1}\nonumber\\
   &=\frac{2/N(1-\gamma)}{4/N+n(\gamma+1)}. \label{Eq: E_n}
\end{align}
We have
\begin{align*}
   & \left |\E\!\left[\gamma_{n+1}\tilde U^1_{n+1}\middle|\F_n\right] \right| 
   \\& = \gamma_{n+1}\Bigg| g_1(X_n)-g_1(\tilde X_n)+g_2(X_n,D_n,E_n)\\
   &-(\gamma+1)\left ( \frac{\gamma}{\gamma+1} -D_n +  \frac{\gamma}{\gamma+1} -E_n  \right ) \Bigg|\\
    &\leq \gamma_{n+1}(k_1+\gamma+1) \left( \left |D_n-\frac{\gamma}{\gamma+1}\right |+ \left |E_n-\frac{\gamma}{\gamma+1}\right |\right)\\
    &+ \gamma_{n+1}k_2 \left(\frac{1}{\gamma}\left |D_n-\frac{\gamma}{\gamma+1}\right |+ \left |E_n-\frac{\gamma}{\gamma+1}\right |\right) \\
    &\leq C \gamma_{n}^2
\end{align*}
for some constant $C>0$, where the first equality follows from $\E[U^1_{n+1}|\F_n]=0$ and $X_n,\tilde X_n,D_n,E_n \in \F_n$, the second inequality follows from the Lipschitz continuity of $g_1$ (with Lipschitz constant $k_1$) and $G^1$ (with Lipschitz constant $k_2$), and the definition of $g_2$ \eqref{Eq: g_2}, and the last inequality is a consequence of \eqref{Eq: D_n}-\eqref{Eq: E_n} and $\gamma_{n+1}=\frac{1}{(4/N)+n(\gamma+1)}$. 

The second moment of the noise $\tilde U^1_{n+1}$ satisfies
\begin{align*}
   \E\!\left[(\tilde U_{n+1}^1)^2\middle|\F_n\right]
   &\leq 2 \Bigg (g_1(X_n)-g_1(\tilde X_n)+g_2(X_n,D_n,E_n)\\
   &-(\gamma+1)\left ( \frac{\gamma}{\gamma+1} -D_n +  \frac{\gamma}{\gamma+1} -E_n  \right ) \!\Bigg )^{\!2} \\
   & + 2 \E\!\left[( U_{n+1}^1)^2\middle|\F_n\right]
   \\ & \leq C_1 \left ( \left |D_n-\frac{\gamma}{\gamma+1}\right |^2+ \left |E_n-\frac{\gamma}{\gamma+1}\right |^2 \right ) \\
   &+ 2 \E\!\left[(U_{n+1}^1)^2\middle|\F_n\right] 
\end{align*}
for some constant $C_1>0$. By \eqref{Eq: U1}, we obtain 
\begin{align*}
   & \E\!\left[( U_{n+1}^1)^2\middle|\F_n\right]  \\ 
   & = \gamma^2 \E\!\left[(\Psi_n - (1-\Phi)(1-p_n^0)-\Phi p_n^1 )^2\middle|\F_n\right] \\ 
    & = \gamma^2 \E\!\left[((1\!-\!\Phi) \Psi_n^0 + \Phi \Psi_n^1 - (1\!-\!\Phi)(1-p_n^0)-\Phi p_n^1 )^2\middle|\F_n\right] \\
    & = \gamma^2(1-\Phi)^2 \E\!\left[ (\Psi_n^0 - (1-p_n^0))^2\middle|\F_n \right] \\ 
    & + \gamma^2\Phi^2 \E\!\left[(\Psi_n^1 - p_n^1)^2\middle|\F_n \right] \\ 
    & = \gamma^2(1-\Phi) p_n^0(1-p_n^0) +\gamma^2 \Phi p_n^1(1-p_n^1) \\
    & = \gamma^2(1-\Phi) \frac{X_n}{X_n+Y_n} (1-p_n^0) \\
    &+\gamma^2 \Phi p_n^1\left (1-\frac{Z_n}{1-X_n-Y_n}\right ) \\
    & \leq \gamma^2 (\gamma+1)(1-\Phi) X_n + \gamma^2 (\gamma+1) \Phi (1-X_n-Y_n-Z_n) \\
    & = \gamma^2(\gamma+1) \Bigg  ( (1-\Phi) X_n + \Phi \bigg ( \frac{X_n}{\gamma} + \frac{1}{\gamma} \left( \frac{\gamma}{\gamma+1}-D_n \right) \\ &+ \frac{\gamma}{\gamma+1} - E_n \bigg ) \Bigg)  \leq C_2 |\tilde X_n  |
\end{align*}
for some constant $C_2>0$, where the second equality follows from \eqref{Eq: psi_n}, the third equality follows from the conditional independence of $\Psi_n^0$ and $\Psi_n^1$, the fourth equality follows from  $N\Psi_n^0 \sim \bin(N,1-p_n^0)$  and $N\Psi_n^1 \sim \bin(N,p_n^1)$, and  the sixth inequality follows from part 1 of Lemma \ref{lem: domains}. Hence,  
$$ \E\!\left[(\tilde U_{n+1}^1)^2\middle|\F_n\right] \leq C_3 |\tilde X_n|$$
for some constant $C_3>0$.

Using similar arguments to those used in Proposition \ref{Prop: Non-existence of Almost Sure Limit}, one can show that $\Psi_n$ does not converge to $0$ or $1$. Hence, $nX_n \to \infty$ as $n\to \infty$. Following \eqref{Eq: D_n} and \eqref{Eq: E_n}, $\tilde X_n - X_n = O(1/n)$. This implies that  $n \tilde X_n \to \infty$ as $n\to \infty$. We have $g_1(x)x>0$ in the neighborhood of $0$.
Moreover, using the definition of $M$ in Lemma \ref{lem: domains} and \eqref{Eq: g_1}, one can show that $g_1(x)^2 = O(x)$.
Hence, by \cite[Theorem 3]{renlund2011recursive}, we deduce that $\P( \tilde X_n \to 0)=0$. We deduce that $\P(X_n \to 0)=0$, which proves  the case $\gamma \geq  \max\left( \frac{\Phi}{1-\Phi},\frac{1-\Phi}{\Phi}\right)$. 

\section{Proof of Lemma  \ref{lemma: individual convergence}}
\label{Proof of Lemma individual convergence}

\underline{Part 1:}
Consider the case $\phi_v = 1$. The proof for the $\phi_v = 0$ case is analogous. Define the stochastic process
\[
z_{n}=\sum_{k=0}^{n-1} \left(\psi_{v,k} - p_{k}^1 \right).
\]
The process $z_n$ is $\F_n$-adapted. Since $\E[\psi_{v,n}|\F_{n}]=p_{n}^1$, we obtain that $\E[z_{n+1}|\F_{n}]=z_{n}$. Thus, $z_n$ is an $\F$-martingale. Furthermore, $|z_n-z_{n-1}| \leq 2$. Using the \emph{Azuma-Hoeffding inequality} \cite{azuma1967weighted}, we deduce that for all $\epsilon>0$, 
\[
\P \left( \left|\frac{1}{n}z_{n} \right|> \epsilon \right) \leq 2\exp \left( -\frac{\epsilon^2 n^2}{2 \sum_{k=1}^n 2^2} \right) = 2\exp \left( -\frac{\epsilon^2 n}{8} \right).
\]
By the Borel-Cantelli lemma \cite[Lemma 2.5]{Cinlar2011}, we obtain that the event $A_n=\big\{\left|\frac{1}{n}z_{n} \right| > \epsilon\big\}$ occurs finitely often. Thus, $\lim\limits_{n \to \infty} \frac{1}{n}z_{n} =0$. Observe that $\frac{1}{n} \sum_{k=0}^{n-1}p_{k}^1$ converges to $p_{\infty}^1$ by the dominated convergence theorem (or Ces\`{a}ro summation). This proves the result.

\underline{Part 2:}
We have 
\begin{align*}
  \hat \Psi_n &= \frac{1}{n} \sum_{k=0}^{n-1} \frac{1}{N}\sum_{v\in V}\psi_{v,k} = \frac{1}{N} \sum_{v\in V}\frac{1}{n}  \sum_{k=0}^{n-1} \psi_{v,k}\\
  &=\frac{1}{N} \sum_{v\in V,\phi_v=0}\frac{1}{n}  \sum_{k=0}^{n-1} \psi_{v,k}+\frac{1}{N} \sum_{v\in V,\phi_v=1}\frac{1}{n}  \sum_{k=0}^{n-1} \psi_{v,k}.  
\end{align*} 
Using part 1, we obtain
\begin{align*}
    \lim\limits_{n \to \infty}\frac{1}{N} \sum_{v\in V,\phi_v=0}\frac{1}{n}  \sum_{k=0}^{n-1} \psi_{v,k} = \frac{|V^0|}{N} & = (1-\Phi)(1-p_\infty^0),\\
    \lim\limits_{n \to \infty}\frac{1}{N} \sum_{v\in V,\phi_v=1}\frac{1}{n}  \sum_{k=0}^{n-1} \psi_{v,k} = \frac{|V^1|}{N} & = \Phi p_\infty^1,
\end{align*}
where $|V^j|$ is the number of agents that have inherent opinion $j$.
This proves the second part.

\section{Proof of Theorem \ref{Thm: global estimation}}
\label{Proof of Theorem global estimation}

\underline{Part 1:} The first part follows from part 2 of Lemma  \ref{lemma: individual convergence}. 

 \underline{Parts 2 and 3:} To prove the second part, note that $g_\infty = \frac{1}{\gamma+1}$ if and only if  $\hat \Psi_\infty=0$. By Lemma \ref{lemma: individual convergence}, if $\hat \Psi_\infty=0$, then $(p_\infty^0,p_\infty^1)=(1,0)$. Then, by Theorem \ref{Thm: convergence}, $\gamma \leq  \frac{1-\Phi}{\Phi}$, which is equivalent to $\Phi \leq  \frac{1}{\gamma+1}$. The third part is shown similarly. 

\underline{Part 4:} We have $g_\infty \in \big(\frac{1}{\gamma+1},\frac{\gamma}{\gamma+1}\big)$ if and only if $\Psi_\infty \in (0,1)$.  Lemma \ref{lemma: individual convergence} implies that $(p_\infty^0,p_\infty^1) \notin \{(0,1),(1,0)\}$. So, Theorem \ref{Thm: convergence} implies that $(p_\infty^0,p_\infty^1)=\big( \frac{\gamma(\gamma(1-\Phi)-\Phi)}{(1-\Phi)(\gamma^2-1)},\frac{\gamma(\gamma\Phi-(1-\Phi))}{\Phi(\gamma^2-1)} \big)$. Therefore, 
$$\hat \Psi_\infty= (1-\Phi) (1-p_\infty^0) + \Phi p_\infty^1=\frac{\gamma \Phi - (1-\Phi)}{\gamma-1} , $$
where the second equality follows from \eqref{Eq: Estimator_general}. This completes the proof.

\section*{Acknowledgment}

The authors are grateful to Govind Ramnarayan for stimulating discussions about the problem formulation.

\balance

\bibliographystyle{IEEEtran}
\bibliography{refs}

\end{document}